\documentclass[preprint,journal]{vgtc}       

\ifpdf
  \pdfoutput=1\relax                   
  \usepackage{graphicx}                
  \DeclareGraphicsExtensions{.pdf,.png,.jpg,.jpeg} 
\else
  \usepackage{graphicx}                
  \DeclareGraphicsExtensions{.eps}     
\fi%

\graphicspath{{figures/}{pictures/}{images/}{./}} 

\usepackage{microtype}                 
\PassOptionsToPackage{warn}{textcomp}  
\usepackage{textcomp}                  
\usepackage{mathptmx}                  
\usepackage{times}                     
\usepackage{cite}                      
\usepackage{tabu}                      
\usepackage{booktabs}                  
\usepackage{comment}
\usepackage{tcolorbox}
\usepackage{makecell}
\usepackage{amsmath}

\usepackage[english]{babel} 
\usepackage{blindtext}
\usepackage{pifont}
\usepackage{utfsym}
\usepackage{multirow}
\usepackage{xcolor}
\usepackage{wasysym}

\DeclareRobustCommand{\NL}{%
  \begingroup\normalfont
  \includegraphics[height=1.3\fontcharht\font`\B]{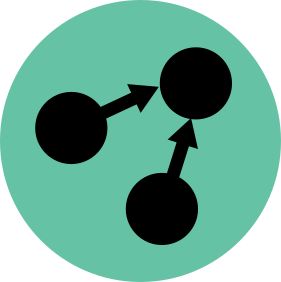}%
  \endgroup
}
\DeclareRobustCommand{\AM}{%
  \begingroup\normalfont
  \includegraphics[height=1.3\fontcharht\font`\B]{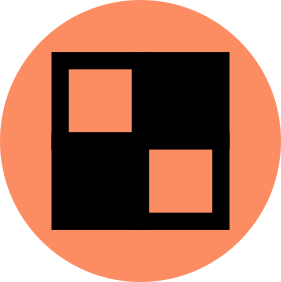}%
  \endgroup
}
\DeclareRobustCommand{\BP}{%
  \begingroup\normalfont
  \includegraphics[height=1.3\fontcharht\font`\B]{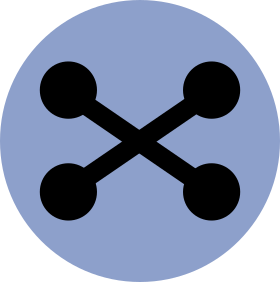}%
  \endgroup
}

\definecolor{mscolor}{rgb}{0,0,0.7}
\definecolor{dwcolor}{rgb}{0,0.7,0.7}
\definecolor{macolor}{rgb}{0.7,0,0.7}
\definecolor{kacolor}{rgb}{0.5,0.7,0}
\definecolor{nscolor}{rgb}{0.5,0,0.5}
\definecolor{kkcolor}{rgb}{0.8,0.3,0.8}

\newcommand\partiallyConf{\textcolor{black}{\ding{51}}{\textsuperscript{\textcolor{black}{\kern-0.5em\small\ding{55}}}}}

\newcommand\mycenter[1]{\par\vspace{-0.2em}
\begin{tcolorbox}[colframe=white, boxrule=0mm,top=0mm, bottom=0mm,enlarge top by=0mm,
enlarge bottom by=0mm]
\centering
#1
\end{tcolorbox}
}

\newcommand{\varB}[1]{{\operatorname{\mathit{#1}}}}

\onlineid{1541}

\vgtccategory{Research}
\vgtcpapertype{Evaluation}

\title{Comparative Evaluation of Bipartite, Node-Link, and Matrix-Based Network Representations}

\author{Moataz Abdelaal, Nathan D. Schiele, Katrin Angerbauer, \textit{Student Member, IEEE}, Kuno Kurzhals, \\Michael Sedlmair, \textit{Member, IEEE}, and Daniel Weiskopf, \textit{Member, IEEE}}
\authorfooter{
\item
 Moataz Abdelaal, Katrin Angerbauer, Kuno Kurzhals, Michael Sedlmair, and Daniel Weiskopf are with University of Stuttgart. E-mail: lastname.firstname@visus.uni-stuttgart.de.
\item
 Nathan D. Schiele is with Leiden University. E-mail: n.d.schiele@liacs.leidenuniv.nl.

}

\shortauthortitle{Abdelaal \MakeLowercase{\textit{et al.}}: Comparative Evaluation of Bipartite, Node-Link, and Matrix-Based Network Representations}

\abstract{This work investigates and compares the performance of node-link diagrams, adjacency matrices, and bipartite layouts for visualizing networks. In a crowd-sourced user study (n = 150), we measure the task accuracy and completion time of the three representations for different network classes and properties. In contrast to the literature, which covers mostly topology-based tasks (e.g., path finding) in small datasets, we mainly focus on overview tasks for large and directed networks. We consider three overview tasks on networks with 500 nodes: (T1)~network class identification, (T2)~cluster detection, and (T3)~network density estimation, and two detailed tasks: (T4)~node in-degree vs.\ out-degree and (T5)~representation mapping, on networks with 50 and 20 nodes, respectively. Our results show that bipartite layouts are beneficial for revealing the overall network structure, while adjacency matrices are most reliable across the different tasks.

} 


\CCScatlist{ 
 \CCScat{K.6.1}{Management of Computing and Information Systems}%
{Project and People Management}{Life Cycle};
 \CCScat{K.7.m}{The Computing Profession}{Miscellaneous}{Ethics}
}

\teaser{
  \centering
  \includegraphics[width=0.98\linewidth]{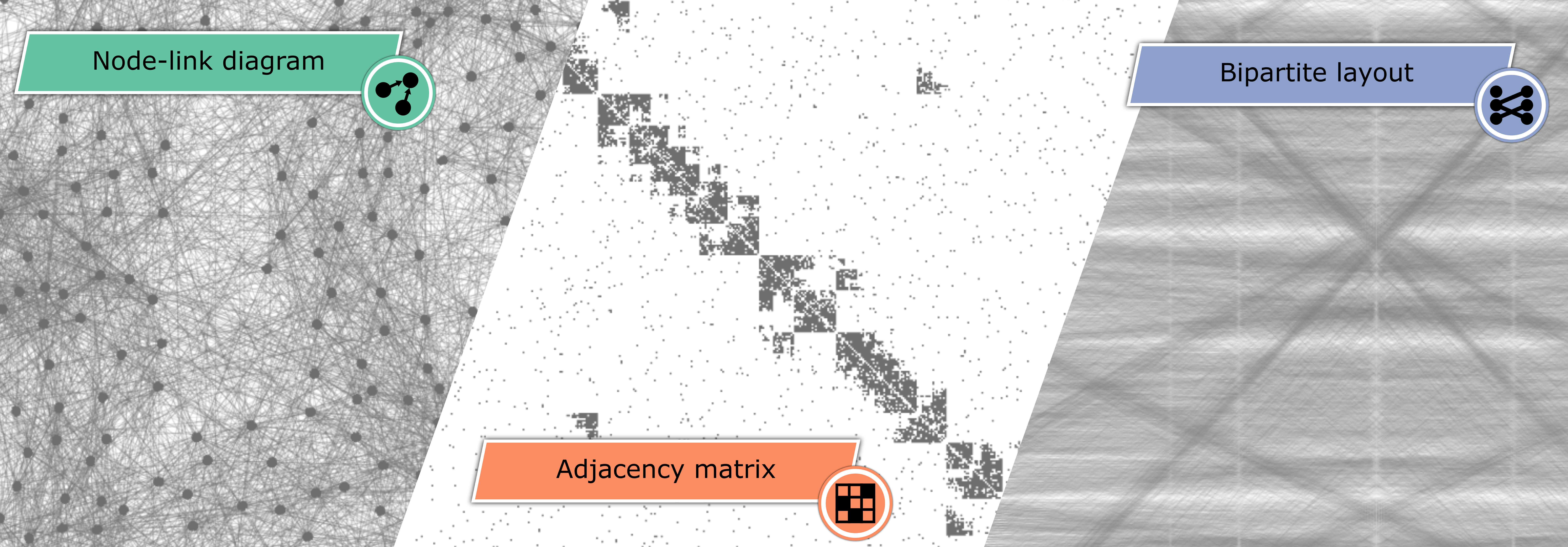}
  \caption{Three visualization techniques for the representation of large networks. Node-link diagrams and adjacency matrices are common in visualization. Bipartite layouts have been proposed as an alternative for solving different tasks. We compare all three techniques with respect to different network properties and tasks.}
  \label{fig:teaser}
}

\vgtcinsertpkg
\usepackage{subcaption}
\begin{document}


\firstsection{Introduction}

\maketitle

Node-link diagrams (\textbf{NL}) \textbf{\NL} and  adjacency matrices (\textbf{AM}) {\AM} are two of the most common visualizations for  static networks (Figure~\ref{fig:teaser}). \textbf{\NL} is easy to understand but suffers from visual clutter when the size and density of networks increase.
\textbf{\AM} is a visualization free of overdrawing and is highly scalable with respect to network density. However, spatial properties are neglected, impairing path-related tasks~\cite{ Ghoniem2004, Ghoniem2005, Keller2006} (e.g., following a path through a network). To address this problem, several techniques have been proposed, either by combining the visual variables of both representations~\cite{Henry2007NodeTrix, Henry2007MatLink, Sansen2015Edge} or by introducing a whole new representation altogether~\cite{Hlawatsch2014Adjacency, Yoghourdjian2018 ,jin2021trammelgraph}. One of those is the bipartite layout (\textbf{BP}) \BP.

In contrast to the aforementioned techniques, \textbf{\BP} was originally introduced to visualize bipartite graphs~\cite{asratian1998bipartite}, i.e., graphs that have two sets of non-adjacent vertices. 
In recent years, the layout has been proposed as an alternative visual representation of static and dynamic networks~\cite{BurchParallel2011,beck2014state}. The layout depicts networks by connecting nodes on two vertical axes (Figure~\ref{fig:network_BP}). 
Due to its compact design, the layout is often used to \textit{compare} multiple instances of the same network, which is an essential analysis task in dynamic network visualization. 
The technique shares some characteristics with \textbf{\NL} and \textbf{\AM}. It uses the same visual variables as \textbf{\NL} to encode nodes and links. Similar to \textbf{\AM}, it relies on vertex ordering techniques~\cite{LiivSeriation2010,Behrisch2016Matrix,feketeReorderJS2015} to reveal the network structure.
In contrast to \textbf{\AM}, the axes in \textbf{\BP} are parallel and not orthogonal to each other. This makes it more scalable than \textbf{\AM} with respect to the network size~\cite{Burch2017ThousandGraphs,Abdelaal2020TEP}. 
To our knowledge, there is no formal evaluation of \textbf{\BP} thus far.
Despite the increasing number of empirical user evaluations in network visualizations in the last two decades\cite{BurchTheState2021}, the capabilities of \textbf{\BP} for static network visualization remain largely unknown, especially in comparison to \textbf{\NL} and \textbf{\AM}.

In the same vein, Yoghourdjian et al.~\cite{YOGHOURDJIAN2018264} identified multiple aspects of network visualization that have not been investigated by empirical user evaluation. For example, the majority of studies (80\%) were conducted on small-size networks (100 nodes or fewer), and studies that considered large-size networks used network abstraction or aggregation methods for the evaluation~\cite{YOGHOURDJIAN2018264}. While the basic measures of network complexity, such as size and density, are important evaluation parameters, they cannot be viewed independently from the underlying network class or model.
Only few studies~\cite{Yoghourdjian2018} considered network class as an important factor in the evaluation and assessed how it might influence the performance of the participants. 
When it comes to network directionality, 68\% of the studies were based on undirected networks. Regarding user tasks, the most common tasks are the topology-based tasks~\cite{LeeTax2006}, with path finding being the most common; and overview-based tasks are the least common~\cite{YOGHOURDJIAN2018264}. 

In our work, we fill some of the aforementioned gaps by evaluating \textbf{\BP} for network visualization and comparing it to the two traditional approaches \textbf{\NL} and \textbf{\AM}, with respect to multiple network classes, properties, and analysis tasks. Our work extends the body-of-knowledge of empirical evaluations for network visualization in several ways:
\begin{itemize}
    \itemsep-0.2em 
    \item We consider large \textit{directed} networks with 500 nodes.
    \item We explore the interplay between network class and density. In this regard, we consider four different network classes and three density profiles.
    \item We evaluate five tasks that target different levels of network granularity. Three overview tasks (T1 -- T3) on networks with 500 nodes. Two detailed tasks (T4 and T5) on networks with (50 and 20) nodes, respectively.
    \item We contribute new insights regarding \textbf{\BP} as a method for network visualization and how it compares to \textbf{\NL} and \textbf{\AM}.
\end{itemize}
We conducted a large-scale crowd-sourced user study ($n = 150$) to assess the performance of participants for the different network interpretation tasks.  
Our results provide empirical evidence about the advantages and shortcomings of the three techniques under different conditions. Based on the results, we derive guidelines for the use of the three techniques depending on the task and data characteristics.
\section{Related Work}
Related work comprises a multitude of studies investigating node-link and adjacency matrix visualizations. We focus on studies comparing both techniques in terms of task performance. We distinguish ourselves from  previous work with respect to: (1) network properties, with a particular focus on  network size; (2) the inclusion of network class as a variable in the evaluation; and (3) the tasks, with a particular focus on network overview and comparison tasks.
Furthermore, we discuss the use of bipartite graph layouts in the literature and how it was applied for network visualization.

\subsection{Evaluation of Node-Link and Adjacency Matrix}
In the last two decades, there has been an increasing number of empirical user evaluations in network visualization. For a detailed and thorough review, we refer the readers to the surveys of Yoghourdjian et al.~\cite{YOGHOURDJIAN2018264} and Burch et al.~\cite{BurchTheState2021}. An early evaluation of \textbf{\NL} was conducted by Purchase~\cite{PURCHASE1998647}, where they asked the participants to perform three tasks, such as \textit{finding the shortest path between two given nodes} on eight different layouts of {\NL}. They found no statistical difference in performance, except for one layout. The finding is interesting as it shows that graph aesthetic criteria had little influence on task performance.
Ghoniem et al.~\cite{Ghoniem2004,Ghoniem2005} conducted a seminal study in which they compared  \textbf{\AM} and \textbf{\NL} using synthetically generated networks on a number of simple and abstract tasks, such as \textit{finding the most connected node}, or \textit{finding a link between two nodes}. The main result of the work is that \textbf{\AM} is useful for visualizing dense networks, whereas \textbf{\NL} is better suited for path-related tasks. These findings were confirmed later by Keller et al.~\cite{Keller2006} and Henry and Fekete~\cite{Henry2007MatLink} on real-world datasets. 
Beradi et al.~\cite{Berardi2013efficacy} compared \textbf{\NL} and \textbf{\AM} in the context of intelligence tasks such as \textit{identifying the potential leaders
within the network} or \textit{identifying network clusters}. In their evaluation, \textbf{\NL} performed better with respect to accuracy and response time. Hlawatsch et al.~\cite{Hlawatsch2014Adjacency} compared both techniques for visualizing weighted, directed, and dynamic networks. Participants were faster and made fewer errors in \textbf{\AM} when they had to compare edge weights across several points in time.

While  previous work evaluated 
 basic measures of complexity such as size and density, more recent studies tried to leverage crowdsourcing evaluation to compare \textbf{\NL} and \textbf{\AM} across a multitude of network tasks. For example, Okoe et al. recruited 557  online users in~\cite{Okoe2018} and 864 in~\cite{Okoe2019} to evaluate 14 different tasks on two real-world datasets. Similarly, Ren et al.~\cite{Ren2019} recruited 600 participants to evaluate how people understand social networks by measuring their performance in 16 different tasks. Nobre et al.~\cite{Nobre2020} compared \textbf{\NL} and \textbf{\AM} in the context of  multivariate networks. They evaluated 16 tasks with a special focus on attribute-based tasks. Except for the work of Okoe et al.~\cite{Okoe2018, Okoe2019}, most of the previous studies evaluated networks with 100 nodes or less. Additionally, most of the evaluated tasks are low-level (i.e., topology- or attribute-based tasks)~\cite{LeeTax2006}, except for the task \textit{``How many clusters are there in the visualization?''} in the work of Okoe et al.~\cite{Okoe2018, Okoe2019}. In contrast to previous work, we consider larger networks with 500 nodes. Additionally, we focus our evaluation more on network overview and comparison tasks. 

There are a few prior studies that considered network overview and comparison tasks. For example, Alper et al.~\cite{Alper2013} evaluated network comparison tasks on both \textbf{\NL} and \textbf{\AM}, using brain connectivity networks. Similarly, Jin et al.~\cite{jin2021trammelgraph} considered ``overall similarity” as one of the tasks in their evaluation. 
Our work builds upon that by evaluating large networks and investigating different  network classes in the evaluation. We position ourselves close to the work done by Yoghourdjian et al.~\cite{Yoghourdjian2018}, where they compared \textbf{\NL} and \textbf{\AM} using large networks with thousand nodes across different network classes. In contrast to their work, we investigate the interplay between the network class and network density by 
including the density as an additional parameter in the evaluation. Additionally, while Yoghourdjian et al.~evaluated one overview task (i.e., identifying similarity), we extend that procedure by evaluating three overview tasks and two detailed tasks for directed networks. Furthermore, we leverage a large number of participants by conducting a crowdsourcing user study with 150 participants. 
Finally, we evaluate \textbf{\BP} layout as a method for visualizing directed networks and compare it against the two standard techniques \textbf{\NL} and \textbf{\AM}, which has not been done to date.

\subsection{Bipartite Graph and Layout}

Bipartite graphs are a special kind of graph where the vertices are partitioned into two disjoint sets. These graphs are often found in biological and biochemical reaction networks~\cite{asratian1998bipartite}. While there is a substantial body of work on visualizing bipartite graph structures~\cite{misue2006drawing, schulz2008visual, Xu2016, Pezzotti2018}, we focus on using bipartite layouts for visualizing single-mode networks (i.e., one set of vertices). Inspired by parallel coordinates plots, Burch et al.\ proposed Parallel Edge Splatting~\cite{BurchParallel2011}, a technique for visualizing dynamic networks where they used a bipartite layout as the underlying visual representation (\autoref{fig:network_BP}). To encode the time dimension, the individual networks from different timepoints are juxtaposed next to each other in a small multiples fashion. To reduce visual clutter, the edges are splatted by computing a pixel density map. Several subsequent techniques were introduced aiming to increase the scalability of parallel edge splatting with respect to the number of timepoints, either by applying a time sliding window~\cite{Beck2012Rapid}, interleaving~\cite{Burch2017ThousandGraphs}, stacking~\cite{AbdelaalVMV2018}, or by partially drawing the edges~\cite{Abdelaal2020TEP}. All this work showed that bipartite layouts can be beneficial for dynamic network visualization. The 1D arrangement of vertices allows the layout to be scalable with respect to the network size and number of timepoints. Furthermore, with proper vertex ordering, the layout can reveal several network structures~\cite{BurchParallel2011} and is therefore beneficial for network overview and comparison tasks. 
In previous work, evaluations were typically done by showing use cases on a few real-world datasets. While this is a valid evaluation method~\cite{Lam2012}, in our work, we complement the prior work by conducting an empirical user evaluation to assess the capability of bipartite layout for static network visualization and compare it against the traditional network representations.

\section{Network Representations}\label{sec:network_representation}
We investigate three visualization techniques for large and directed graphs. We consider undirected graphs a sub-case of the directed ones. In this section, we briefly introduce the \textbf{\BP} layout as it is less commonly used for visualizing networks and compare it against the baseline techniques, \textbf{\NL} and \textbf{\AM}.

\begin{figure*}
\centering
\begin{subfigure}[t]{0.3\textwidth}
\centering
\includegraphics[height=3.3cm]{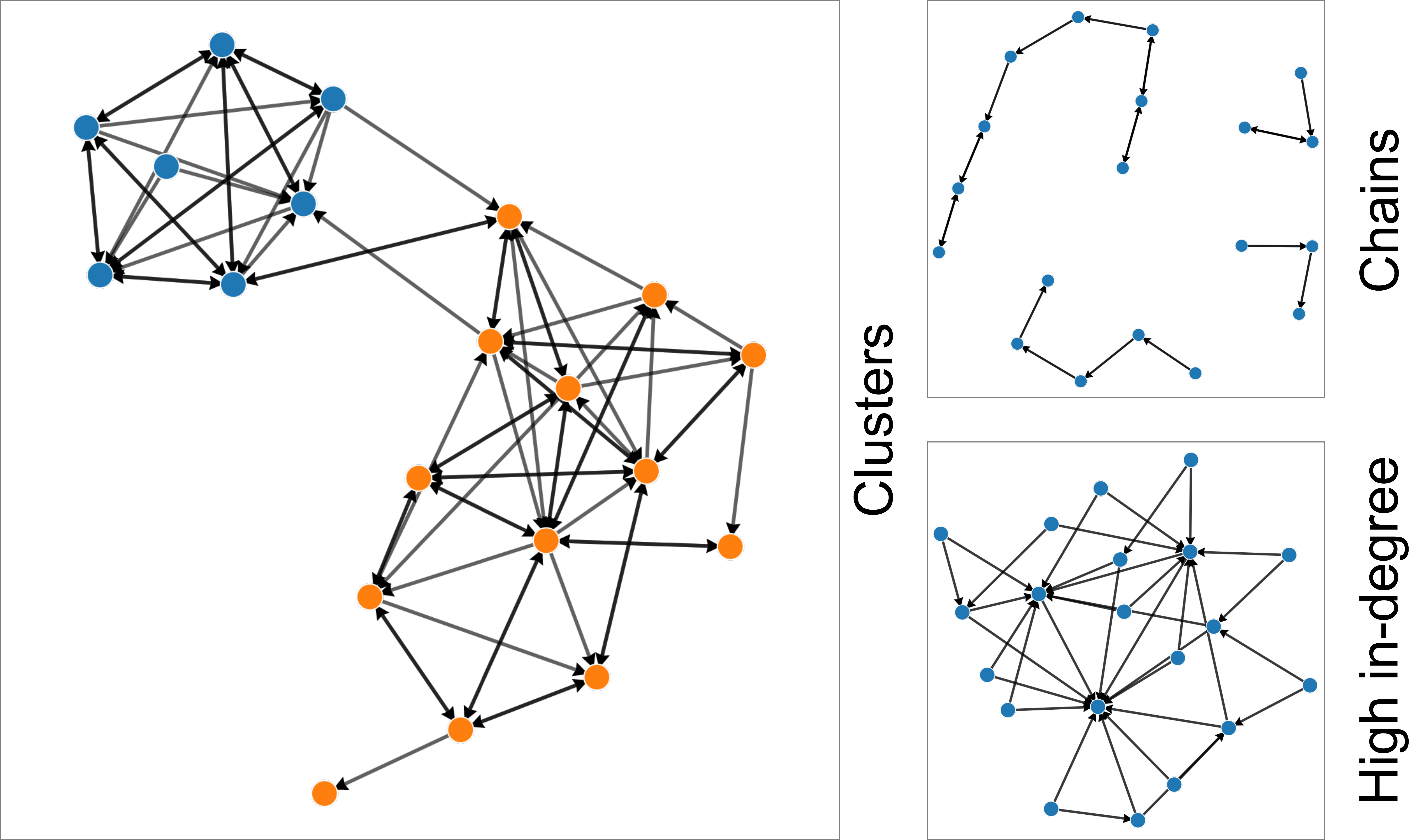}
\subcaption{Node-link diagram \NL}\label{fig:network_NL}
\end{subfigure}
\hfill
\begin{subfigure}[t]{0.3\textwidth}
\centering
\includegraphics[height=3.3cm]{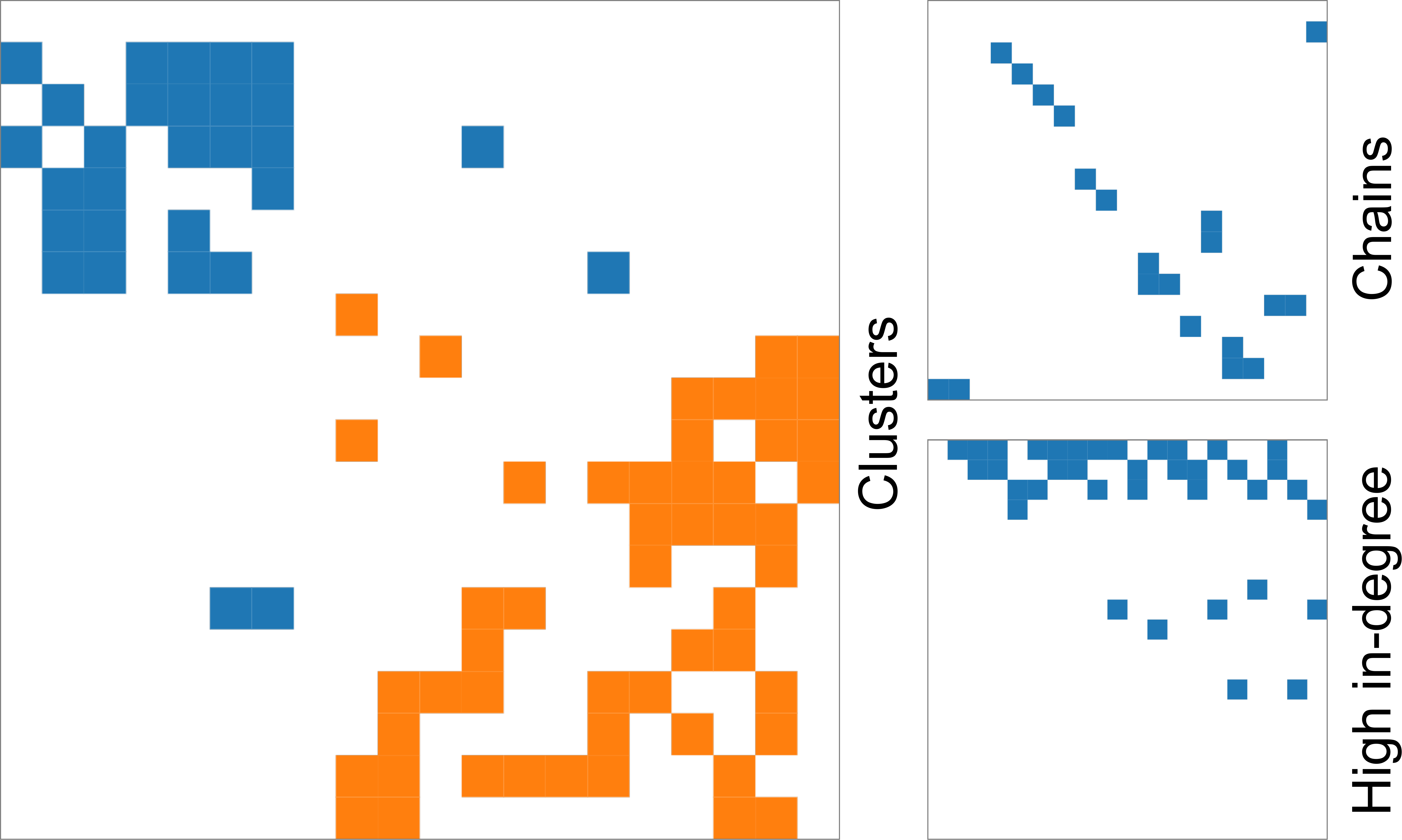}
\subcaption{Adjacency matrix \AM}\label{fig:network_AM}
\end{subfigure}
\hfill
\begin{subfigure}[t]{0.3\textwidth}
\centering
\includegraphics[height=3.3cm]{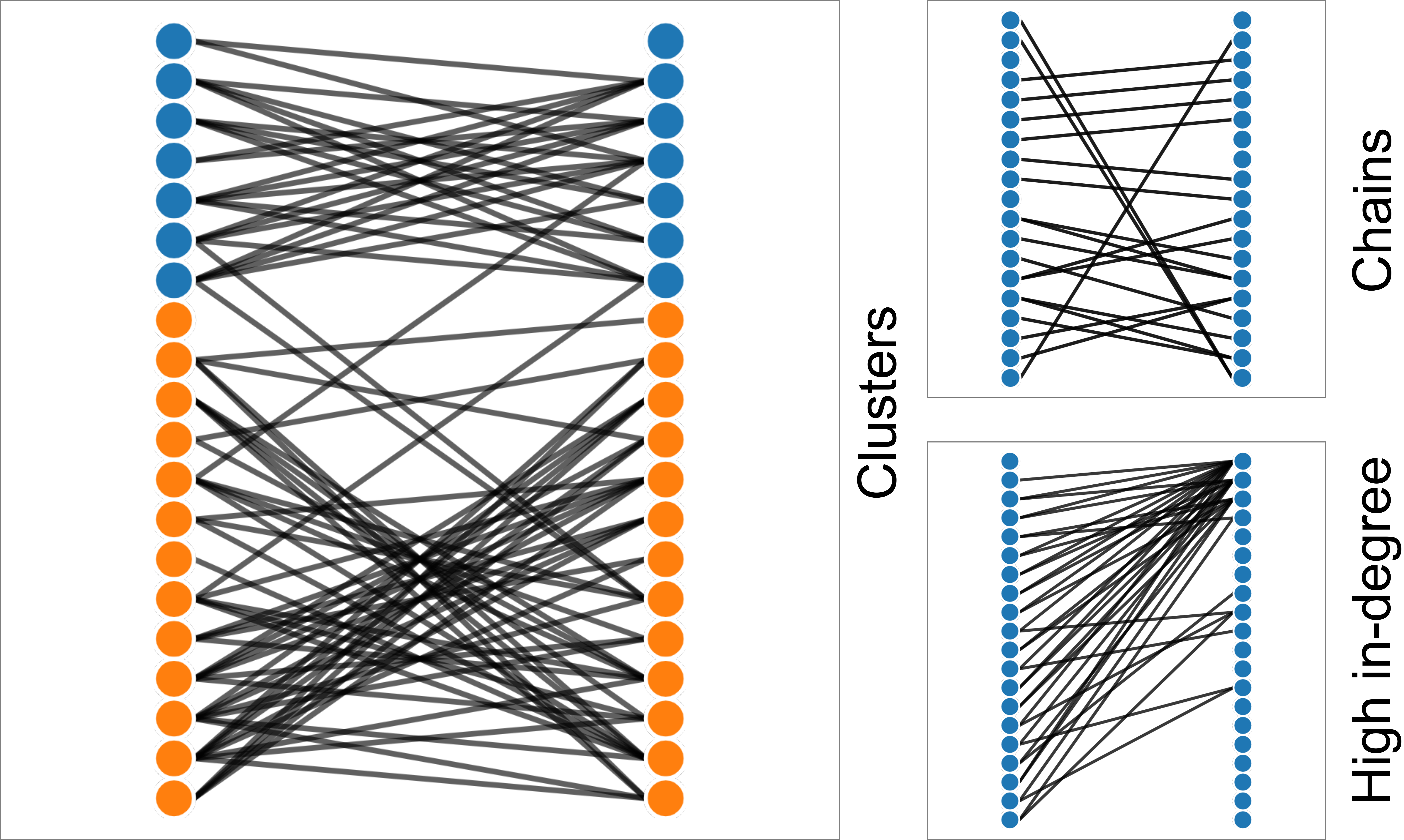}
\subcaption{Bipartite layout \BP}\label{fig:network_BP}
\end{subfigure}

\caption{The three network representations investigated in the study. \textbf{\NL} depicts nodes by dots connected by directional arrows. \textbf{\AM} shows incoming edges row-wise and outgoing ones (column-wise). In \textbf{\BP}, edges go from the nodes on the left axis to duplicates of nodes on the right axis. Visual patterns of clusters, chains, and nodes with high in-degree are apparent in all techniques.}\vspace{-2ex}
\label{fig:network_all}
\end{figure*}

The \textbf{\BP} layout consists of two vertical, parallel axes. The left axis corresponds to the source nodes, while the right axis corresponds to the target nodes. To depict a network, the nodes are replicated and placed on both axes in the same order, and the links are shown by drawing lines connecting the source and target nodes. Therefore, in \textbf{\BP}, the link direction always read from left to right (\autoref{fig:network_BP}). In comparison to \textbf{\NL} and \textbf{\AM}, \textbf{\BP} has a limited scalability with respect to the number of edges, due to the limited drawing space. To mitigate this problem, edge rendering~\cite{BurchParallel2011,Abdelaal2020TEP}~or bundling techniques~\cite{Palmas2014Palmas,Lhuillier2017State} could be applied to improve the visibility of links in cluttered areas. Similar to \textbf{\AM}, \textbf{\BP} relies on vertex ordering algorithms~\cite{Behrisch2016Matrix, LiivSeriation2010}, not only to reveal the network structures, but also to reduce visual clutter. \textbf{\BP} combines some of the features of \textbf{\AM} and \textbf{\NL}. On the one hand, it encodes the connectivity information the same way as \textbf{\NL} does. That is, by drawing lines connecting the source and target nodes. 
On the other hand, similar to \textbf{\AM}, the nodes are replicated and positioned on two axes representing the source and target axes. Having a separate drawing space for network nodes allows \textbf{\BP} and \textbf{\AM} to avoid the problem of overlapping nodes existing in \textbf{\NL}. Therefore, one could hypothesize that \textbf{\BP}, similar to \textbf{\AM}, would be better suited than \textbf{\NL} for tasks that are based on node attributes, such as node lookup or identifying node degree. Additionally, depicting each node twice---once as a source node and once as a target node---could make \textbf{\AM} and \textbf{\BP} better suited than \textbf{\NL} for visualizing directed networks. Despite being inherently built for visualizing directed networks, \textbf{\BP} and \textbf{\AM} can also be used to visualize undirected networks as well. That is, by transforming them into directed ones (i.e., adding the edges in both directions).

Due to the replication of nodes on two separate axes, both \textbf{\AM} and \textbf{\BP} are not well suited for path finding tasks. Both rely on memorizing node labels to solve the task, which might be challenging without the aid of interaction techniques~\cite{ShenyPath2007}, if the path is more than just one hop between the source and target nodes. In contrast, in \textbf{\NL}, one does not need to read the node labels to find a path between two nodes because the path information is visually encoded by both the topological and geometrical properties of the nodes.
In contrast to \textbf{\AM}, the node axes in \textbf{\BP} are rather parallel than orthogonal to each other. Such an arrangement makes \textbf{\BP} more scalable than \textbf{\AM} with respect to network size. However, such scalability comes at the cost of restricting the drawing space to only the area between the two axes, making \textbf{\BP} the least scalable among the three representations with respect to network density.

\autoref{fig:network_all} shows three structural properties of real-world networks and their appearance in each representation. The cluster structure is easy to spot in \textbf{\NL}, due to the two-dimensional spatial layout (\autoref{fig:network_NL}). This pattern forms a rectangular shape along the diagonal (\autoref{fig:network_AM}) in an ordered \textbf{\AM}  or as an hourglass-like shape (\autoref{fig:network_BP}) in an ordered \textbf{\BP}. Sparse, small-world networks exhibit chain-like structures as seen in~\autoref{fig:network_all}(a), at the top-right corner. This structure is translated into parallel lines in \textbf{\BP} or entries along the diagonal in \textbf{\AM}.
Nodes with high in-degree are also common in networks. These nodes form fan-like shapes in \textbf{\BP} or appear as filled rows/columns in \textbf{\AM}. In this example, we use hierarchical clustering to order both \textbf{\AM} and \textbf{\BP} (Section~\ref{subsec:stimuli_data}).

\section{Methods}

Many of the design choices we made throughout the study are based on a qualitative data study~\cite{sedlmair2012taxonomy} that we performed at an early phase of the project. During this phase, we investigated different network properties, classes, and analysis tasks. We also experimented with different ways of rendering networks on each of the aforementioned representations, the use of color, and interactivity. This section details our methodological approach and explains the design choices we made.

\begin{figure*}[t]
    \centering
    \includegraphics[width=0.8\textwidth]{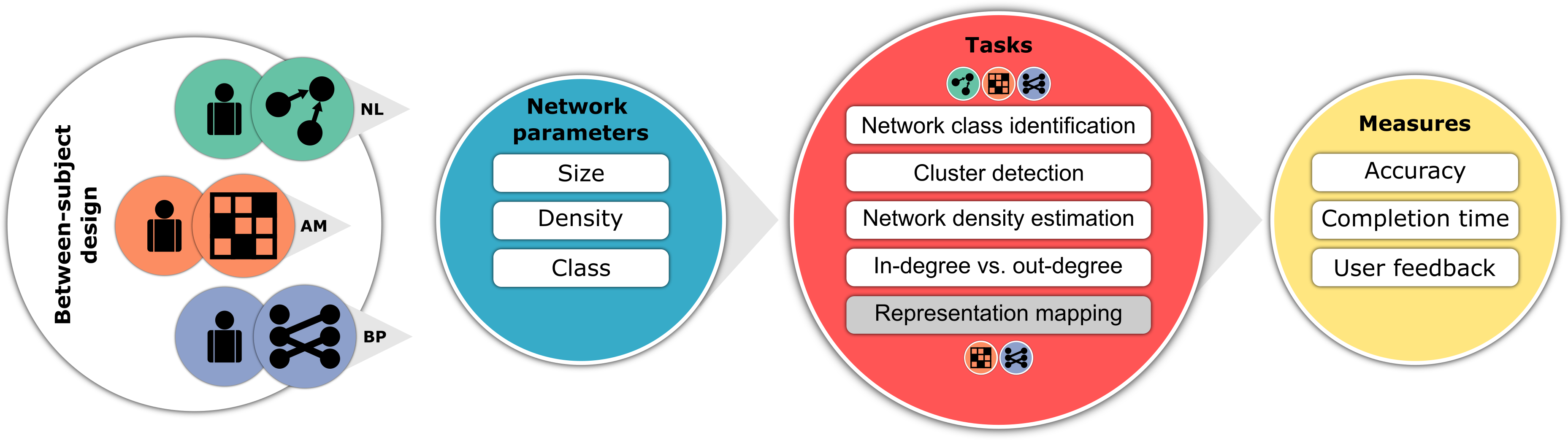}
    \caption{Study design and variables. We presented the visualizations in a between-subject design with a variation in network parameters. All participants performed a set of tasks with one visualization technique. We measured their performance and user feedback for evaluation.}
    \label{fig:study_procedure}\vspace{-2ex}
\end{figure*}

\subsection{Network Parameters}
\label{sec:network_param_sec}
\begin{figure*}[ht]
    \centering
    \includegraphics[width=0.8\textwidth]{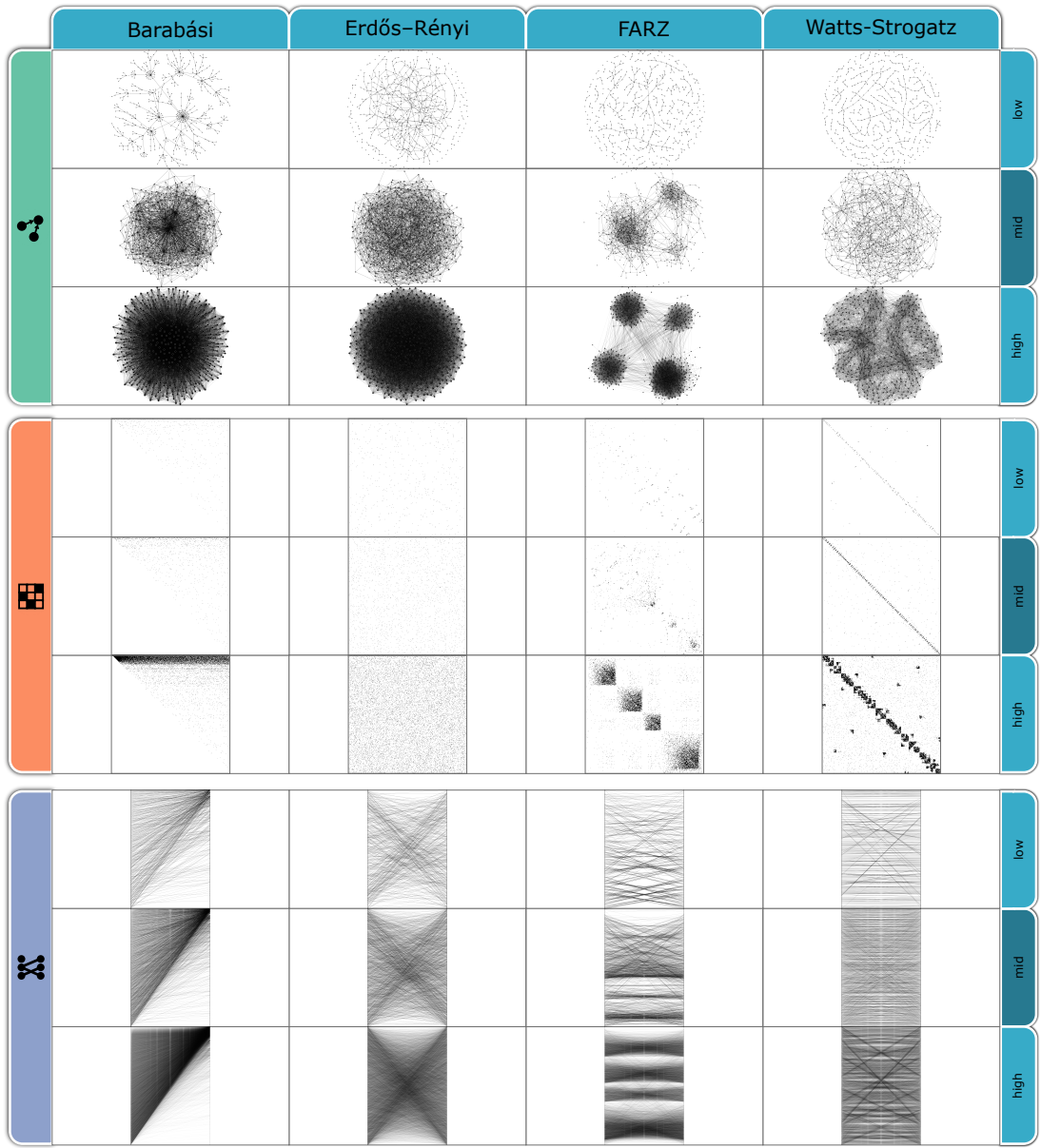}
    \caption{Overview of different network classes presented in the study. For the three visualizations \textbf{\NL}, \textbf{\AM}, and \textbf{\BP}, we presented visual stimuli of each class with different densities. Some tasks become more difficult with decreasing density of the data (e.g., detecting clusters in FARZ).}\vspace{-2ex}
    \label{fig:vis_table}
\end{figure*}

We visualize different networks with respect to density, size, and network class. An overview of how the stimuli change under varying parameters is displayed in \autoref{fig:vis_table}.

\paragraph{Density Profile}
Network density has a strong impact on the visualization. 
As described in the previous section, while \textbf{\AM} scales very well with dense networks, \textbf{\NL} and \textbf{\BP} suffer from overdrawing problems and therefore have limited scalability. Hence, it is relevant to our study to evaluate the different representations under different densities. We adopt a linear density definition with respect to edges and vertices $d = |E|/|V|$, which is  considered to be a good descriptor of densities found in real-world networks~\cite{Melancon2006}. In our study, we aim to evaluate six different densities $d =\{1, 2, 4, 8, 16, 32\}$.
To avoid a large number of trials during the study, we group density values into \emph{density profiles} such that we have three profiles $d_{profile} = \{d_{low}, d_{mid}, d_{high}\}$
, where $d_{low} =\{1, 2\}, d_{mid} =\{4, 8\}$, and $d_{high} =\{16, 32\}$.

\paragraph{Network Size}
Most previous studies used networks with 100 nodes or less~\cite{YOGHOURDJIAN2018264}. In our experiment, we consider networks with approx. 500 nodes for evaluating the overview tasks. Such size is comparable to large networks used in some of the previous studies~\cite{Okoe2018, Okoe2019}. Yoghourdjian et al.~\cite{Yoghourdjian2018} used networks with 1000 nodes. However, since our study does not involve the use of interaction techniques, including such size entails scaling down the resulting network image to fit the visualization to a standard computer screen. While this is true for all three representations, it could impair \textbf{\AM}, making the links hardly visible with low-density profiles. We opted for approx. 500 nodes to provide a fair comparison with \textbf{\AM}. For detailed tasks, we use networks of sizes 20 and 50, which is consistent with previous studies~\cite{Ren2019, Ghoniem2004}.

\paragraph{Network Class}

In our study, we consider four network classes: Barabási~\cite{Barabasi1999}, Erdős-Rényi~\cite{Erds1959}, Watts-Strogatz~\cite{Watts1998}, and FARZ~\cite{Fagnan2018}. Each class comes with different topological features.
Barabási networks exhibit the \emph{preferential attachment} property, where new nodes tend to get connected to the “popular” nodes in the network (i.e., scale-free networks). We call these nodes “the main hubs”, as they are characterized by a high in-degree property.  
Watts-Strogatz networks exhibit small-world network properties: high clustering coefficient and short average path length.
Erdős-Rényi networks have a low clustering coefficient as a result of the random connectivity between the nodes. 
FARZ networks exhibit modular structures (i.e., community structure), a property found in many real-world datasets~\cite{Danon_2005,Fagnan2018}. 
Our selection of the four classes overlaps with the work of Yoghourdjian et al.\cite{Yoghourdjian2018}, who used Barabási, Erdős-Rényi, Watts-Strogatz, and Clustered-Barabási. The latter is a hybrid version of Barabási to emulate the community structure by loosely connecting two separate Barabási networks. In our case, we use FARZ for that purpose.

\subsection{Tasks and Hypotheses}
\label{sec:tasks}
We selected five tasks for evaluation. Tasks T1 -- T3 are network overview tasks, whereas T4 and T5 are detailed tasks for directed networks. The choice of tasks and the formulation of the hypotheses were based on the literature~\cite{LeeTax2006,AmarTax2005,VehlowGroup2017,Yoghourdjian2018, vanHam2008Perceptual} and our qualitative data study. In this section, we briefly explain each task and state our hypotheses along with the rationale behind them.

\paragraph{\textbf{T1: Network Class Identification}}
In this task, we assess the capability of each representation in depicting different network classes and the ability to maintain the visual characteristics of each class under varying densities.
Since we want to rule out memory as a factor in the experiment, we designed this task as a matching task. During the experiment, we showed the participants one network image at the top and asked them: \textit{To which class does the network belong?} At the bottom of the screen, we showed the participants a reference image for each network class to choose from. In other words, the participants had to find out which reference image is the most similar to the given network image at the top. To account for the density parameter, we paired different density profiles of the given network image with different density profiles of the reference images. Since we had three density profiles, this resulted in six different pairs. In total, there are $4$ (classes) $\times$ $6$ (density profile pairs) = $24$ trials, per visualization condition. Our hypothesis is as follows:

\mycenter{$H_{1}$: \textbf{{\AM} and {\BP} are more accurate than {\NL} for T1.}}

As seen in ~\autoref{fig:vis_table}, \textbf{\AM} and \textbf{\BP} are able to maintain the visual characteristics of each network class irrespective of the density profile. In contrast, the visual characteristics seem to disappear as the density changes in \textbf{\NL}. This can be seen in dense networks, where the hairball generated by Barabási does not differ much from the one generated by Erdős-Rényi. However, this kind of problem also occurs at low density, where it is hard to differentiate FARZ from Watts-Strogatz. Yoghourdjian et al.~\cite{Yoghourdjian2018} arrived at a similar conclusion when they compared \textbf{\AM} against \textbf{\NL} across different network classes.

\paragraph{\textbf{T2: Cluster Detection}}
In this task, we assess the capability of each representation in revealing network clusters under varying densities. We only considered networks generated by the FARZ model (Section~\ref{sec:network_param_sec}). 
In this task, there are two parameters to vary: the density profile $d_{profile}$ 
and the number of clusters $k = \{1, 2, 3, 4, 5, 6, 7, 8\}$. 
For the low-density profile, we excluded density $d = 1$, since the generated networks are too sparse to form any cluster. 
We arranged the $k$ values into four groups with consecutive values. 
In total, there were 3 (density profiles) $\times$ 4 (cluster groups) = 12 different combinations. For each, we sampled two networks, making 24 trials, per visualization. During the experiment, we showed the participants one network image and asked them: \textit{How many clusters can you detect in this network?} Our hypothesis is as follows:

\mycenter{$H_2$: \textbf{{\NL} and {\BP} are more accurate than {\AM} for T2.}}

With respect to this task, Okoe et al.~\cite{Okoe2018, Okoe2019} and Berardi et al.~\cite{Berardi2013efficacy} found no significant difference in accuracy between \textbf{\NL} and \textbf{\AM}. In the study by Nobre et al~\cite{Nobre2020}, \textbf{\AM} performed better than \textbf{\NL}. However, the authors stated that the clusters were difficult to spot in \textbf{\NL}, partially due to the large node sizes that were used to accommodate many of the node attributes (i.e., multivariate network). We expect \textbf{\NL} and \textbf{\BP} to obtain more accurate results than \textbf{\AM}, especially with low-density profiles. As seen in~\autoref{fig:vis_table}, the links are barely visible in \textbf{\AM} when the density profile is low. In contrast, links are more visible in \textbf{\NL} and \textbf{\BP} as a result of overdrawing. One could argue that \textbf{\NL} and \textbf{\BP} amplify the recognition of links in sparse networks, which will lead to a better detection of clusters. 

\paragraph{\textbf{T3: Network Density Estimation}}
In this task, we measure the sensitivity of each representation to changes in network density. We also investigate whether such sensitivity depends on the underlying network density and/or the network class. During the experiment, we showed the participants a pair of network images from the same class side-by-side and asked them: \textit{Which network has more connections (Left or Right)?} In addition to the network class and density profile parameters, we defined an additional parameter $\delta_d$ that accounts for the density difference between the network pairs. We evaluated $\delta_d$ for four values  $\delta_d = \{0.25, 0.5, 0.75, 1.0\}$. This corresponds to one network having $\{25\%, 50\%, 75\%, 100\%\}$  more connections than the other. 
In total, there are $3$ (density profiles) $\times$ $4$ ($\delta_d$ values) $\times$ $4$ (network classes) = $48$ trials, per visualization condition. Our hypothesis is as follows:

\mycenter{$H_3(0)$: \textbf{There is no statistical significance in accuracy or speed between the three representation for T3.}}

Yoghourdjian et al.~\cite{Yoghourdjian2018} and Okoe et al.~\cite{Okoe2018} evaluated the same task and found no statistical significance in accuracy between \textbf{\NL} and \textbf{\AM}. The network properties were $n_{nodes}=20$ in the former, $n_{nodes}=258$ and $n_{links}=1090$ in the latter. There was no information provided regarding the value of $\delta_d$. 
We expect the density difference $\delta_d$ and the density profile $d_{profile}$ to have an influence on the results. With low-density profile and small density difference $\delta_d$, \textbf{\NL} and \textbf{\BP} are expected to be more accurate than \textbf{\AM}, due the low visibility of links in \textbf{\AM}. In contrast, \textbf{\AM} is expected to be more accurate than \textbf{\NL} and \textbf{\BP} in high density and with small density differences $\delta_d$, due to the edge overdrawing in \textbf{\NL} and \textbf{\BP}. Regarding the network class, the density comparison task on \textbf{\AM} and \textbf{\BP} is expected to be more accurate for Erdős-Rényi and Barabási than for FARZ and Watts-Strogatz. As seen in~\autoref{fig:vis_table}, Erdős-Rényi networks are characterized by a uniform distribution of links. This makes the density comparison task 
relatively easy, since one does not have to pay attention to the network details but rather to the overall picture. In Barabási networks, participants are expected to pay attention to the part of the network where the “main hubs” are located (i.e., the top part of the visualization). In the case of FARZ and Watts-Strogatz, the networks are composed of several clusters or sub-communities and, therefore, the comparison task involves inspecting the density within each cluster separately, which makes the task harder, and therefore, participants are expected to be less accurate, especially in \textbf{\NL}. 

\paragraph{\textbf{T4: Node In-degree Vs.\ Out-degree}}
Comparing in- and out-degrees of nodes is one of the most common tasks on a directed network. 
In contrast to the previous tasks, we considered a relatively small network ($n_{nodes}=50$). While we originally planned to evaluate the task on large networks ($n_{nodes}=500$), we found out that solving the task is not realistic without the use of interaction techniques. We avoided including interaction features since they would have introduced another degree of freedom in the experiment, and it would have been quite hard to implement interactive features without favoring one representation over the others. 
In this task, there are two parameters:  $d_{profile}$ and the network class. We excluded the Barabási class as it is trivial to solve. Therefore, there are $3$ (density profiles) $\times$ $3$ (network classes) $\times$ $3$ (repetitions) = $27$ trials, per visualization condition. During the experiment, we showed the participants one network image where one node is randomly highlighted and asked them: \textit{Does the highlighted node have more incoming or outgoing links?} Our hypothesis is as follows:

\mycenter{$H_4$: \textbf{{\AM} is more accurate than {\BP} and {\NL} for T4.}}

\textbf{\NL} and \textbf{\BP} both suffer from overdrawing problems, whereas \textbf{\AM} does not. Nevertheless, since \textbf{\BP} provides two separate axes for source and target nodes, the participants are expected to be more accurate in \textbf{\BP} than in \textbf{\NL}. We expect $d_{profile}$ to influence the results. For mid-density and high-density profiles, \textbf{\NL} would require the participants to visit the two endpoints of each link to determine whether it is bidirectional or not, which is a more laborious and error-prone process than for \textbf{\BP} and \textbf{\AM}. In contrast, with low-density profiles, we expect no differences, which is supported by findings from previous work~\cite{Keller2006,Hlawatsch2014Adjacency}. We expect the network class to not have an influence on the results as this is a detailed task that does not take into account the overall structure of the network, which is more amplified by highlighting the node under question. We decided not to use color to encode direction information in \textbf{\NL}, as it would have introduced a bias into the experiment toward one representation. Instead, we reduced the size of the arrows and altered their shape to enhance their recognition in cluttered areas. 

\begin{table*}[t]
\centering
\caption{Summary of tasks and presented stimuli.}
\label{tab:task-table}
\resizebox{\textwidth}{!}{%
\renewcommand{\arraystretch}{1.2}
\begin{tabular}{lccccccccccc|ccc|}
\cline{2-15}
\multicolumn{1}{l|}{\multirow{3}{*}{}} & \multicolumn{1}{c|}{\multirow{3}{*}{$n_{nodes}$}} & \multicolumn{6}{c|}{$d_{profile}$}                                                                                                                                                            & \multicolumn{4}{c|}{\multirow{2}{*}{Network class}}                                               & \multicolumn{3}{c|}{\multirow{2}{*}{Trials}}                 \\ \cline{3-8}
\multicolumn{1}{l|}{}                  & \multicolumn{1}{c|}{}                   & \multicolumn{2}{c|}{low}                                  & \multicolumn{2}{c|}{mid}                                  & \multicolumn{2}{c|}{high}                                 & \multicolumn{4}{c|}{}                                                                             & \multicolumn{3}{c|}{}                                        \\ \cline{3-15} 
\multicolumn{1}{l|}{}                  & \multicolumn{1}{c|}{}                   & \multicolumn{1}{c|}{1}      & \multicolumn{1}{c|}{2}      & \multicolumn{1}{c|}{4}      & \multicolumn{1}{c|}{8}      & \multicolumn{1}{c|}{16}     & \multicolumn{1}{c|}{32}     & \multicolumn{1}{c|}{Barabási} & \multicolumn{1}{c|}{Erdős}  & \multicolumn{1}{c|}{FARZ}  &  Watts  &   \multicolumn{1}{c|} {\NL}   & \multicolumn{1}{c|} {\AM}   & {\BP}   \\ \hline
\multicolumn{1}{|l|}{T1: Class Identification}               & \multicolumn{1}{c|}{500}                & \multicolumn{1}{c|}{\CIRCLE} & \multicolumn{1}{c|}{\CIRCLE} & \multicolumn{1}{c|}{\CIRCLE} & \multicolumn{1}{c|}{\CIRCLE} & \multicolumn{1}{c|}{\CIRCLE} & \multicolumn{1}{c|}{\CIRCLE} & \multicolumn{1}{c|}{\CIRCLE}  & \multicolumn{1}{c|}{\CIRCLE} & \multicolumn{1}{c|}{\CIRCLE} & \CIRCLE & \multicolumn{1}{c|}{24} & \multicolumn{1}{c|}{24} & 24 \\ \hline
\multicolumn{1}{|l|}{T2: Cluster Detection}               & \multicolumn{1}{c|}{500}                & \multicolumn{1}{c|}{}       & \multicolumn{1}{c|}{\CIRCLE} & \multicolumn{1}{c|}{\CIRCLE} & \multicolumn{1}{c|}{\CIRCLE} & \multicolumn{1}{c|}{\CIRCLE} & \multicolumn{1}{c|}{\CIRCLE} & \multicolumn{1}{c|}{}        & \multicolumn{1}{c|}{}       & \multicolumn{1}{c|}{\CIRCLE}       &  & \multicolumn{1}{c|}{24} & \multicolumn{1}{c|}{24} & 24 \\ \hline
\multicolumn{1}{|l|}{T3: Density Estimation}               & \multicolumn{1}{c|}{500}                & \multicolumn{1}{c|}{\CIRCLE} & \multicolumn{1}{c|}{\CIRCLE} & \multicolumn{1}{c|}{\CIRCLE} & \multicolumn{1}{c|}{\CIRCLE} & \multicolumn{1}{c|}{\CIRCLE} & \multicolumn{1}{c|}{\CIRCLE} & \multicolumn{1}{c|}{\CIRCLE}  & \multicolumn{1}{c|}{\CIRCLE} & \multicolumn{1}{c|}{\CIRCLE} & \CIRCLE & \multicolumn{1}{c|}{48} & \multicolumn{1}{c|}{48} & 48 \\ \hline
\multicolumn{1}{|l|}{T4: Node In-degree Vs.\ Out-degree}               & \multicolumn{1}{c|}{50}                 & \multicolumn{1}{c|}{\CIRCLE} & \multicolumn{1}{c|}{\CIRCLE} & \multicolumn{1}{c|}{\CIRCLE} & \multicolumn{1}{c|}{\CIRCLE} & \multicolumn{1}{c|}{\CIRCLE} & \multicolumn{1}{c|}{\CIRCLE} & \multicolumn{1}{c|}{}        & \multicolumn{1}{c|}{\CIRCLE} & \multicolumn{1}{c|}{\CIRCLE} & \CIRCLE & \multicolumn{1}{c|}{27} & \multicolumn{1}{c|}{27} & 27 \\ \hline
\multicolumn{1}{|l|}{T5: Representation Mapping}               & \multicolumn{1}{c|}{20}                 & \multicolumn{1}{c|}{\CIRCLE} & \multicolumn{1}{c|}{}       & \multicolumn{1}{c|}{}       & \multicolumn{1}{c|}{}       & \multicolumn{1}{c|}{}       & \multicolumn{1}{c|}{}       & \multicolumn{1}{c|}{\CIRCLE}  & \multicolumn{1}{c|}{} & \multicolumn{1}{c|}{\CIRCLE} & \CIRCLE & \multicolumn{1}{c|}{-}    & \multicolumn{1}{c|}{12}   & 12   \\ \hline
\multicolumn{12}{l|}{}                                                                                                                                                                                                                                                                                                                                                   & \multicolumn{1}{c|}{123}  & \multicolumn{1}{c|}{135}  & 135  \\ \cline{13-15} 
\end{tabular}%
}
\vspace{-2ex}
\end{table*}

\paragraph{\textbf{T5: Representation Mapping}}
The design of this task was inspired by the work of Yoghourdjian et al.~\cite{Yoghourdjian2018} and Kriglstein et al.~\cite{KriglsteinTransform2018}.
In this task, we want to measure which representation --- \textbf{\AM} or \textbf{\BP} --- is “easier” to map to \textbf{\NL}. We used \textbf{\NL} as a reference representation, since it is known to be an intuitive and easy-to-understand representation of networks. Therefore, we hypothesized that a network representation is easy to understand if it could be mapped “easily” to \textbf{\NL}.
We used small ($n_{nodes} = 20$) and sparse ($d = 1$) networks and varied the network class parameter. We intentionally left out the node labels to force participants into looking at the network structure rather than doing a one-to-one mapping. We also excluded Erdős-Rényi networks since pilot testing showed that they were particularly hard to solve due to the absence of distinctive structure properties, compared to other classes (\autoref{fig:vis_table}).
During the experiment, we showed the participants one network image in \textbf{\BP} or \textbf{\AM} representation at the top and provided them with two \textbf{\NL} diagrams at the bottom and asked them: \textit{Which {\NL} diagram corresponds to the given representation at the top?} In total, there are $3$ (network classes) $\times$ $4$ (repetitions) = $12$ trials, per visualization condition. Our hypothesis is as follows:

\mycenter{$H_5$: \textbf{{\BP} is more accurate than {\AM} for T5.}}

Without node labels, we expect this task to be generally challenging to solve regardless of the underlying visual representation. This was also confirmed during pilot testing. 
However, since \textbf{\NL} and \textbf{\BP} use the same visual variables to encode nodes and links, we hypothesize that it would be easier to solve the task in \textbf{\BP} than \textbf{\AM}. In \textbf{\AM}, participants will have to count the rows and columns to  uniquely identify the nodes which would be more laborious, even with small networks. Out of the three classes, we expect Barabási to be the easiest to solve since it is characterized by the existence of the “main hub” nodes, which would be easier to spot in either representation. 

\subsection{Stimuli Data}
\label{subsec:stimuli_data}
We generated  stimuli data for each task. \autoref{tab:task-table} gives an overview of the different network parameters used in each task. When a task involves multiple repetitions (i.e., T4 and T5), a new network is generated for each repetition. We decided on generating synthetic networks in order to have fine-grained adjustability of network parameters. 
Task T1 revolves around being able to discern separate network classes, which requires relatively objective classes in the study stimulus. Real-world data often exhibit traits of multiple network classes, making it hard to find a ground truth for the task. Additionally, task T2 revolves around finding the number of clusters in the provided network. This requires us to have an objective measure of exactly how many clusters are present in a given network. This is trivial with a generated dataset, as the number of clusters is a generation parameter (i.e., $k$) of the FARZ network model. 
With real-world data, it is often unclear how many clusters are present. We set the $k$ parameter to $4$ for tasks T1 and T3, $1$ for tasks T4 and T5, and $\{1, 2, 3, 4, 5, 6, 7, 8\}$ for task T2. We re-implemented the FARZ model~\cite{Fagnan2018}  
and used the \textit{igraph R package}~\cite{igraph} to generate the data for the other classes. We used the \textit{d3-force}~\cite{d3-force:online} algorithm to layout the \textbf{\NL} diagram. Our experience~\cite{darus-supplemental} shows that \textit{d3-force}
obtains results that are comparable with the best-of-breed layout algorithms such as neato or sfdp. To order the vertices of \textbf{\AM} and \textbf{\BP}, we used agglomerative hierarchical clustering~\cite{hclustfu46:online} with average linkage. To calculate the similarity between two nodes, we use the Jaccard distance based on the sets of direct neighbors. Hierarchical clustering is widely used in the literature and known to provide good results~\cite{wilkinson2012grammar,Behrisch2016Matrix}.

For \textbf{\NL}, we applied the most common technique for graph visualization with solid dots for vertices and straight lines with arrowheads for edges (\autoref{fig:network_NL}). 
For \textbf{\AM}, the columns were used to depict the outgoing links while the rows depicted the incoming ones. In that sense, the link direction always reads top to left (\autoref{fig:network_AM}). The visual stimuli were rendered as static images with no interactivity. We omitted the use of color to focus on the spatial layout of the different techniques. We used the opacity variable in general to increase the visibility of links in \textbf{\NL} and \textbf{\BP}. It was also used to highlight the nodes in task T4. Finally, we downscaled the images by pixel averaging, to make sure they could fit into a browser window.

\subsection{Study Design and Procedure}

We used a between-subject study design with three conditions, one per representation (\autoref{fig:study_procedure}). For a large number of participants and variety of demographics, we conducted an online crowdsourcing user study using Amazon Mechanical Turk (mTurk). 
Similar to prior work~\cite{Radu2014HowtoDisplay, Chapman2014Sets, Okoe2018}, we aimed for 50 participants per condition. 
The order of presented stimuli was counter-balanced
between the conditions. In the case of \textbf{\BP} and \textbf{\AM}, 
participants had to perform five tasks, and only four (excluding T5) in the case of \textbf{\NL}. The order of the tasks was assigned randomly.

At the beginning of the study, participants were presented with a demographics questionnaire. 
After that, participants read a brief tutorial that explained the basics of network visualization and gave an introduction to the visualization being evaluated. Participants could browse through the tutorial back and forth until they felt ready to perform the tasks. Each task started with basic instructions and examples, followed by a few training trials where participants had a chance to see the correct answer. When participants were ready, they performed the actual trials. For each trial, we recorded the participants' answers and response times. After all trials of one task, we asked participants to rate the difficulty of the task on a 5-point Likert scale from “Very Easy” to “Very Difficult” and inform us about the strategy they followed to solve the task. Both questions were required fields. The study had in total 135 trials for \textbf{\AM} and \textbf{\BP} and 123 trials for \textbf{\NL} (\autoref{tab:task-table}).
The study web-application was built using client-server architecture using PHP, JavaScript, and jsPsych~\cite{de2015jspsych}. We refer to the supplemental materials for the source code, data files, and screenshots of the study~\cite{darus-supplemental}.

\subsection{Study Participants and Piloting}
We required participants to be at least 18 years old, speak fluent English, and have a screen-resolution of at least $1920\times1080$ px. The first and second conditions were self-reported. The third was verified automatically using the jsPsych library. To ensure a high quality of responses on mTurk, we set a Human Intelligence Task (HIT) success rate $\ge$ 97\% and a total number of accepted HITs $\ge$ 1000. Furthermore, we added several attention-check trials throughout the study. We filtered out participants who did not pass the attention-check trials or those who took less than 10 minutes to complete the study. In total, we recruited 150 online users after applying our filtering criteria (35\% female). The majority of the participants (44\%) had an age between 31 and 40 years, followed by 19\% above 50 years of age. 65\% of the participants were located in North America, followed by 17\% located in Asia. 

Our pilot testing consisted of two phases. First, we recruited three visualization experts to test-run the study in an informal laboratory setting. The objective of this phase was to verify the clarity of the tasks and the usability of the user interface and to get an initial estimation of the time and the difficulty of each task. During the testing, we encouraged the participants to think aloud and ask questions. In the end, we held an open discussion where we asked the participants for general feedback. In addition to shortening the text, fixing the typos, and fine-tuning the user interface, one of the main results of this phase was to exclude the Erdős-Rényi class from task T5 as participants showed signs of frustration or quickly gave up when they were presented with Erdős-Rényi stimuli compared to other classes. The second piloting phase involved running a test batch of 20 participants on mTurk. The objective of this phase was not only to test the functionality of the study web application but also to ensure the quality of the responses and obtain a more accurate estimation of the length of the study. 
The main outcome of this phase was adding the attention-check trials. Our pilot testing showed that the study lasted for approximately 45 minutes. We compensated the participants with a rate of \$12 per hour.

\section{Results}
\begin{figure}[t]
    \centering
    \includegraphics[width=\columnwidth]{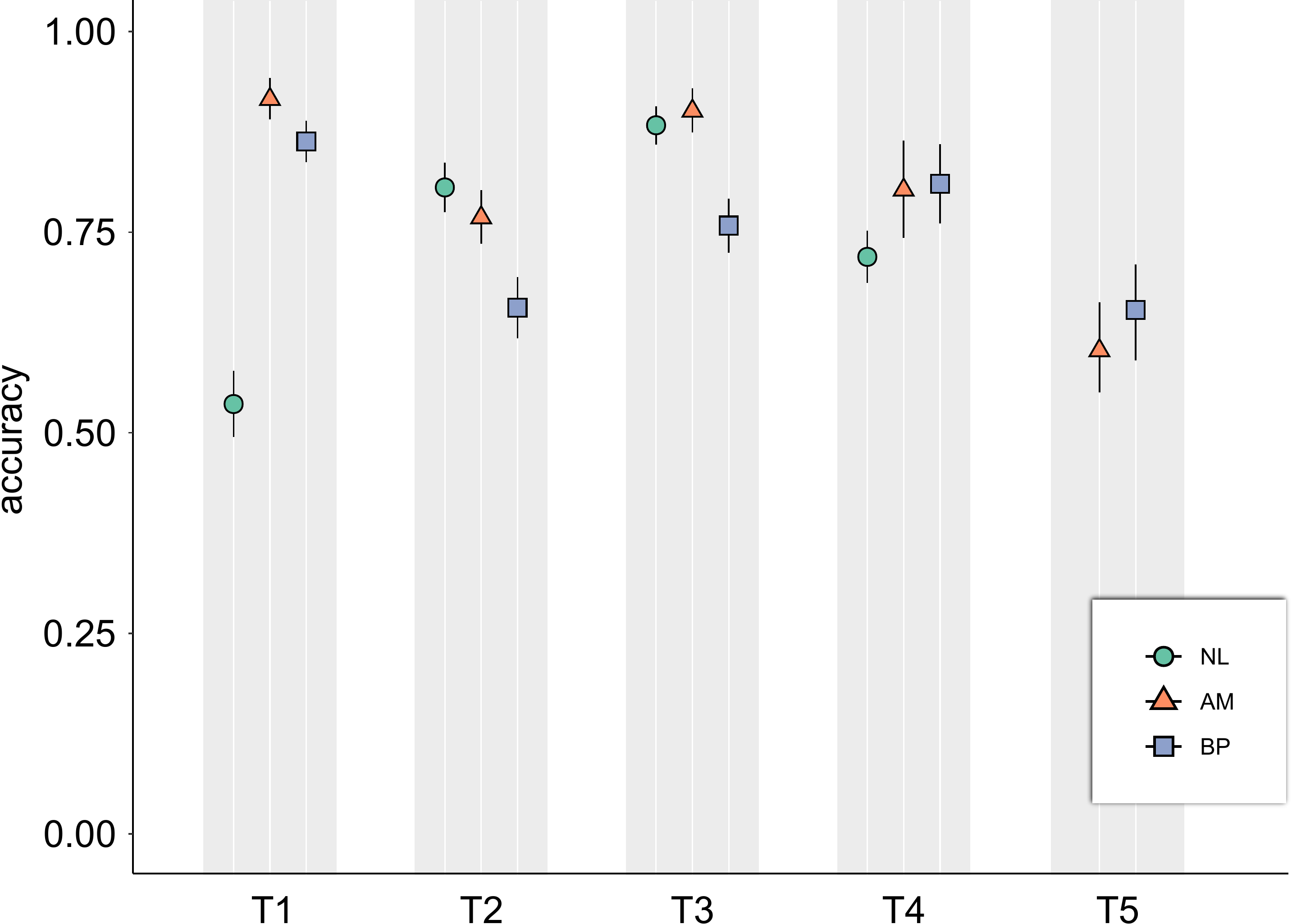}
    \caption{The overall accuracy of tasks T1 -- T5. The shapes represent the means, and the error bars their 95\% CIs.}\vspace{-2ex}
    \label{fig:results-overview}
\end{figure}

To investigate the overall hypothesis, we do statistical testing with Kruskal-Wallis and Wilcoxon Signed Rank tests, as the data was not normally distributed. In addition, we provide 95\% confidence intervals (CIs) to enrich the evaluation with an estimation-based approach. We also use this estimation-based approach for further exploratory investigations of the network parameters, as recommended by Cumming~\cite{cumming2014}. 
As the network parameters were within-subject, we also normalize the between-subject variability~\cite{morey2008confidence,loftus1994using}. Figure~\ref{fig:results-overview} shows an overview of the accuracy of the different representations across the tasks in our study and Table~\ref{tab:statistics} reports the results of the statistical analysis. The raw data and the source code used for statistical analysis are provided in the supplemental materials~\cite{darus-supplemental}.

\def\var_width{0.24}
\newlength{\varheight}
\setlength{\varheight}{4.5cm}
\begin{figure*}[t]
\centering
  \begin{subfigure}[t]{\var_width\textwidth}
      \centering
    \includegraphics[height=\varheight]{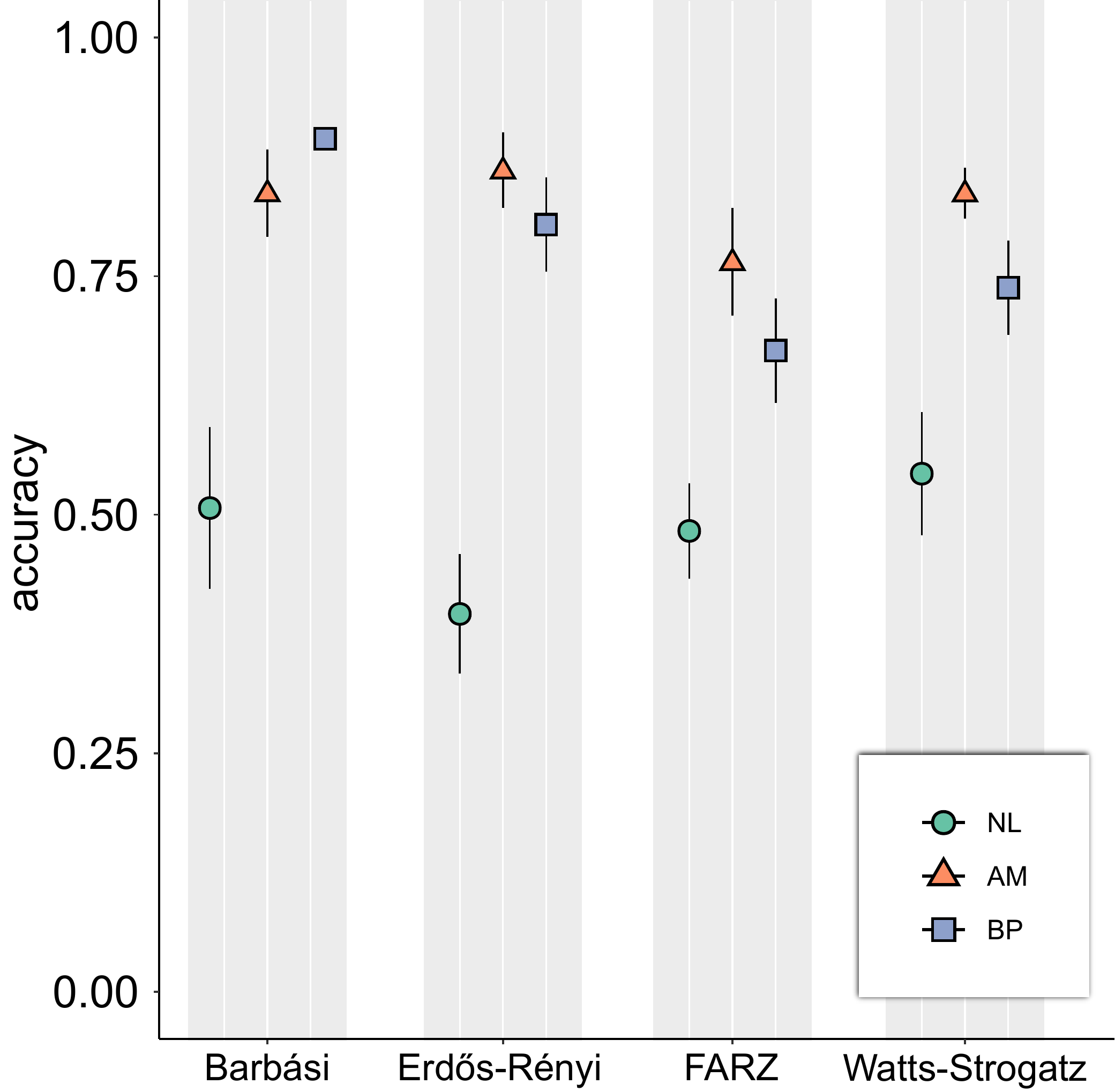}
    \caption{T1: Network Class Parameter}
    \label{fig:t1_class}
  \end{subfigure}
  \hfill
\begin{subfigure}[t]{\var_width\textwidth}
      \centering
    \includegraphics[height=\varheight]{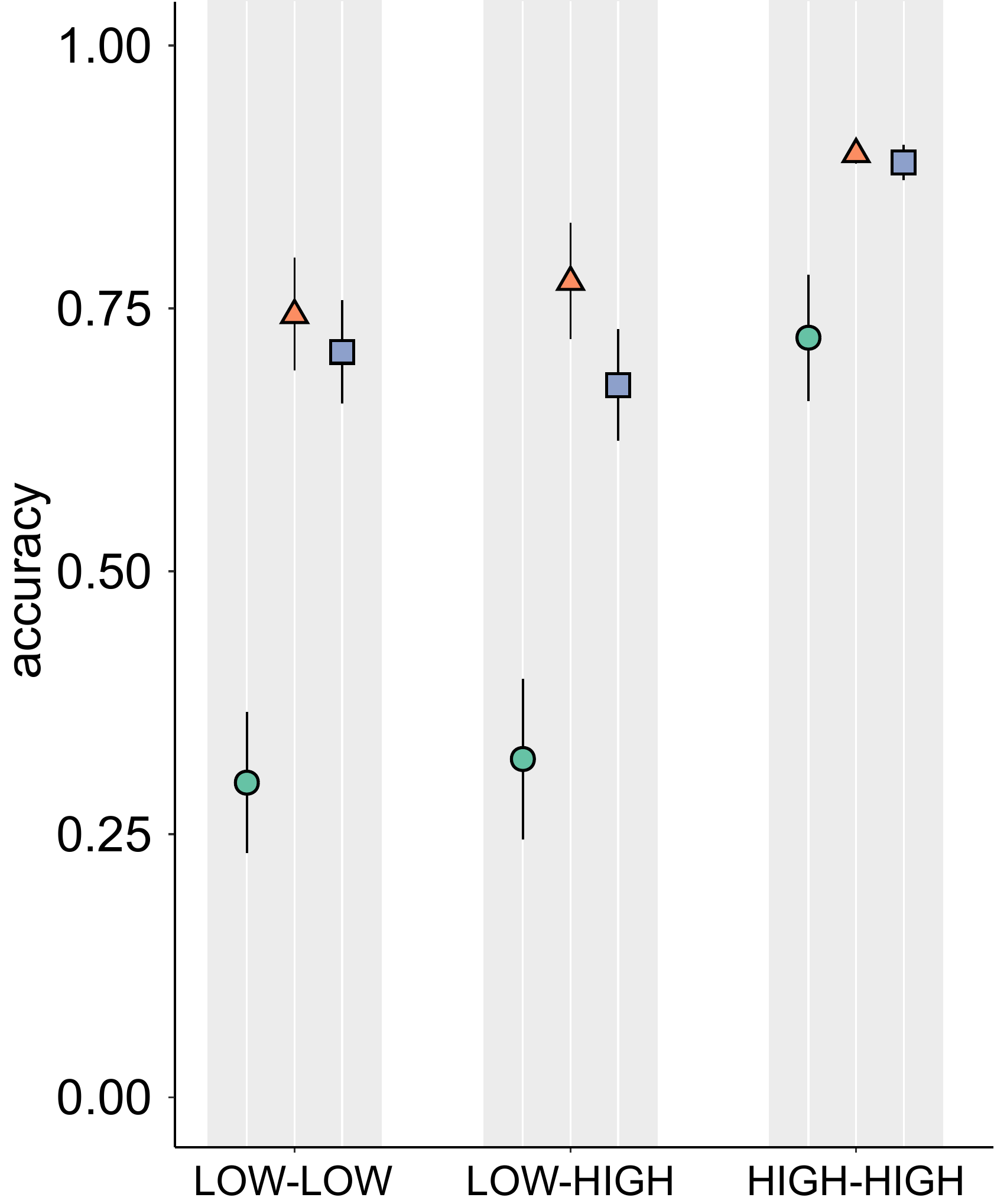}
    \caption{T1: Density Pairing}
    \label{fig:t1_density}
\end{subfigure}
\hfill
\begin{subfigure}[t]{\var_width\textwidth}
      \centering
    \includegraphics[height=\varheight]{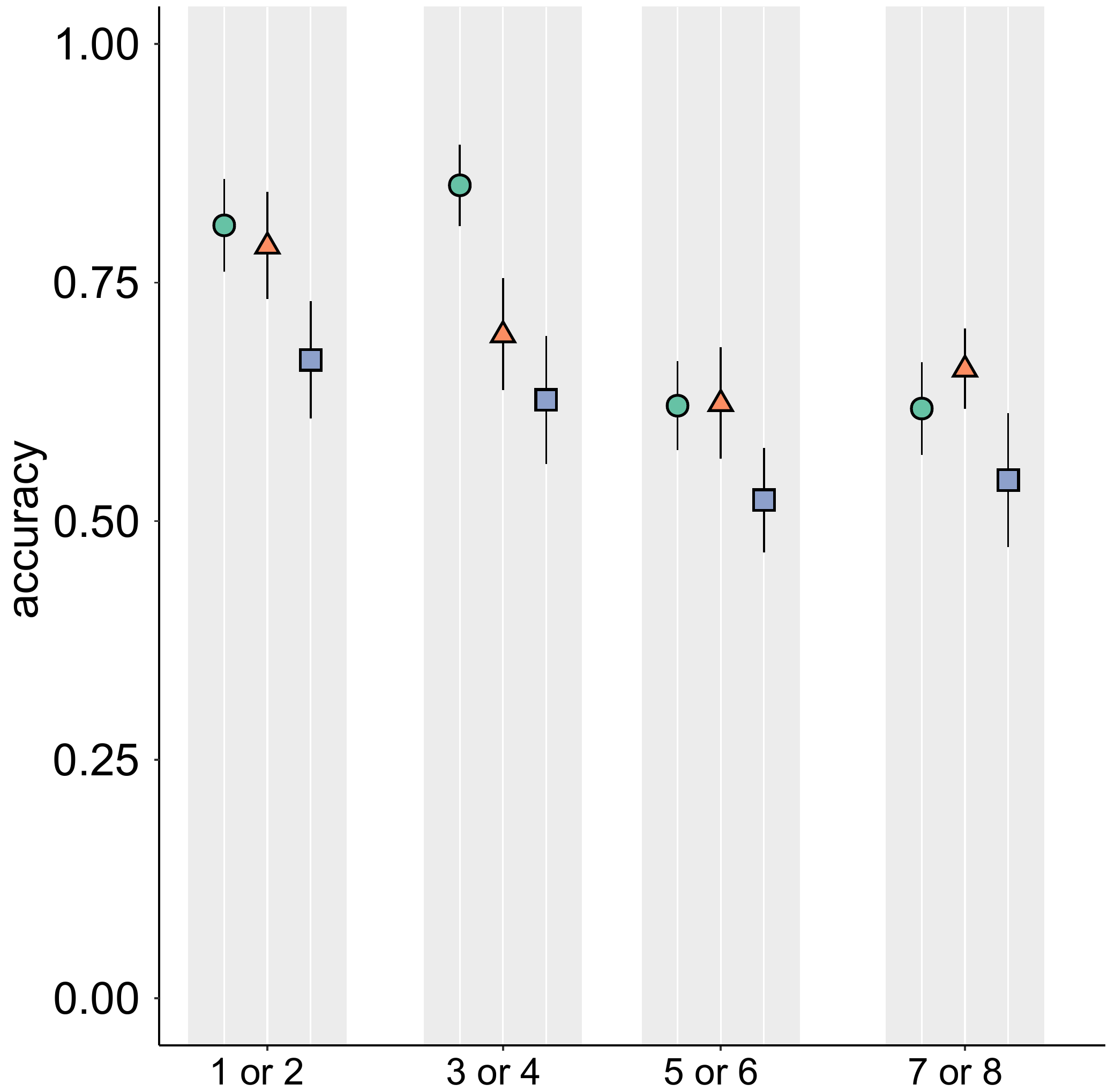}
    \caption{T2: No. of clusters $k$}
    \label{fig:t2_cluster}
\end{subfigure}
\hfill
\begin{subfigure}[t]{\var_width\textwidth}
      \centering
    \includegraphics[height=\varheight]{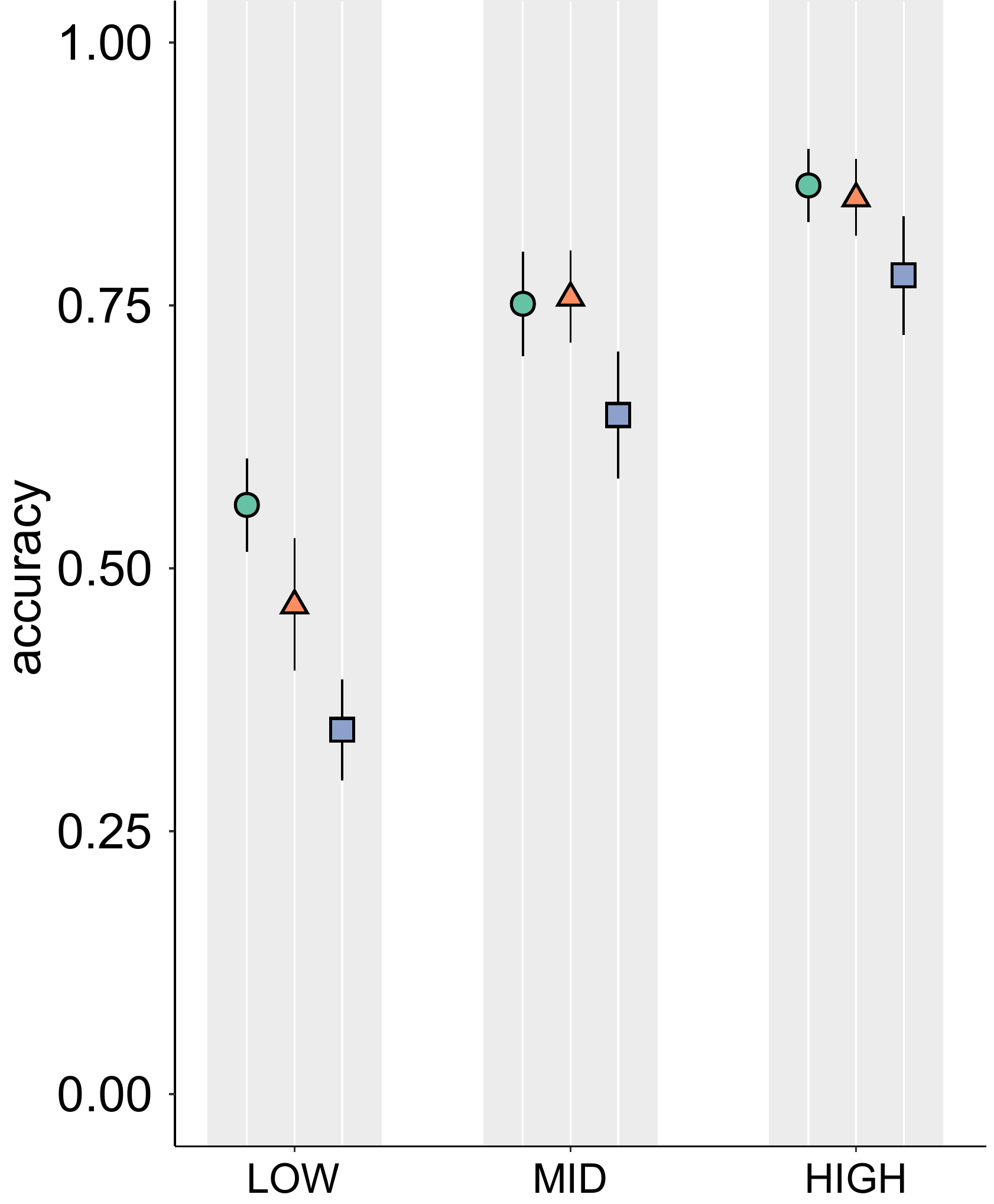}
    \caption{T2: Density Profile}\vspace{2ex}
    \label{fig:t2_density}
\end{subfigure}

\begin{subfigure}[t]{\var_width\textwidth}
      \centering
    \includegraphics[height=\varheight]{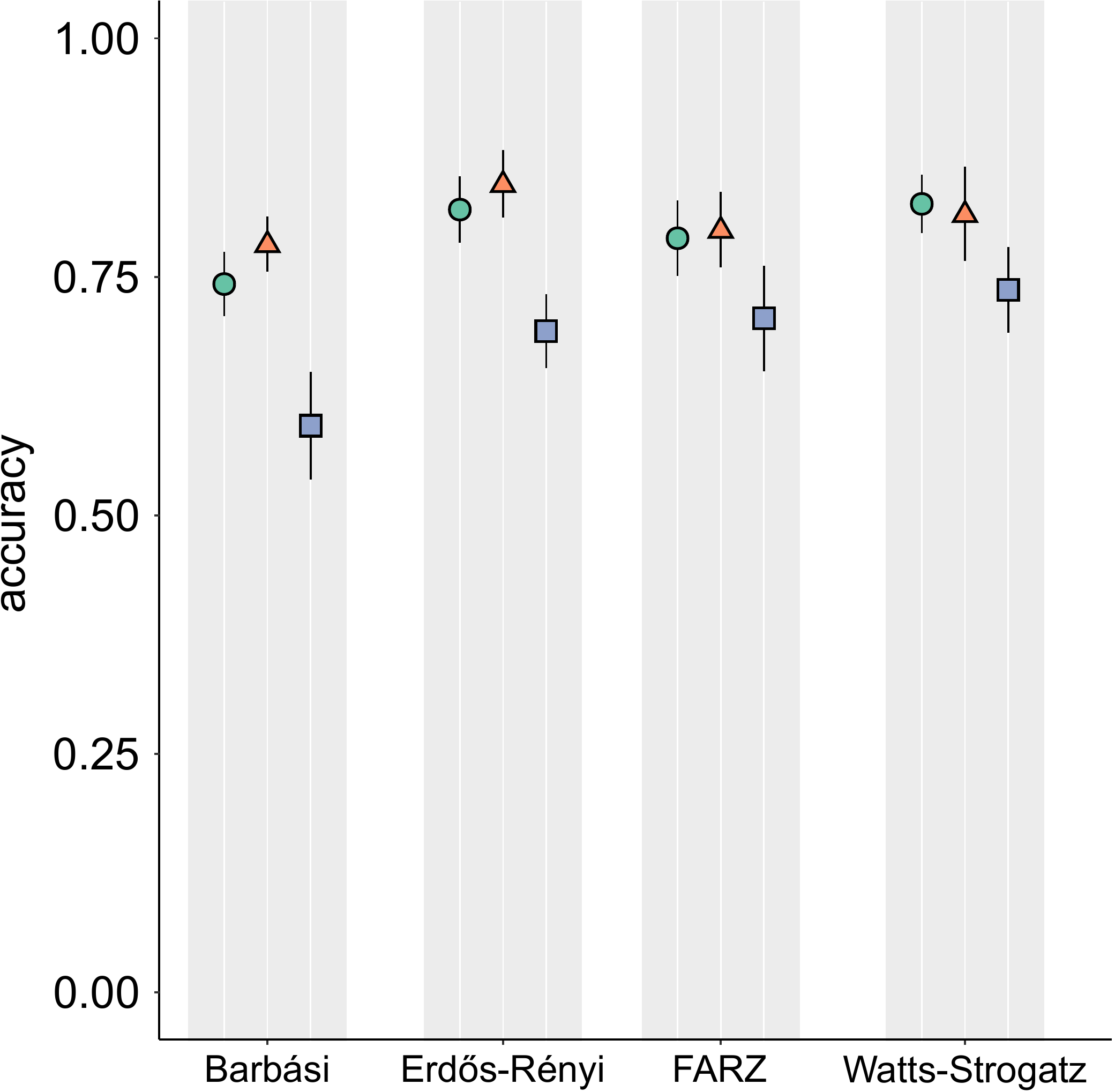}
    \caption{T3: Network Class Parameter}
    \label{fig:t3_cluster}
\end{subfigure}
\hfill
\begin{subfigure}[t]{\var_width\textwidth}
      \centering
    \includegraphics[height=\varheight]{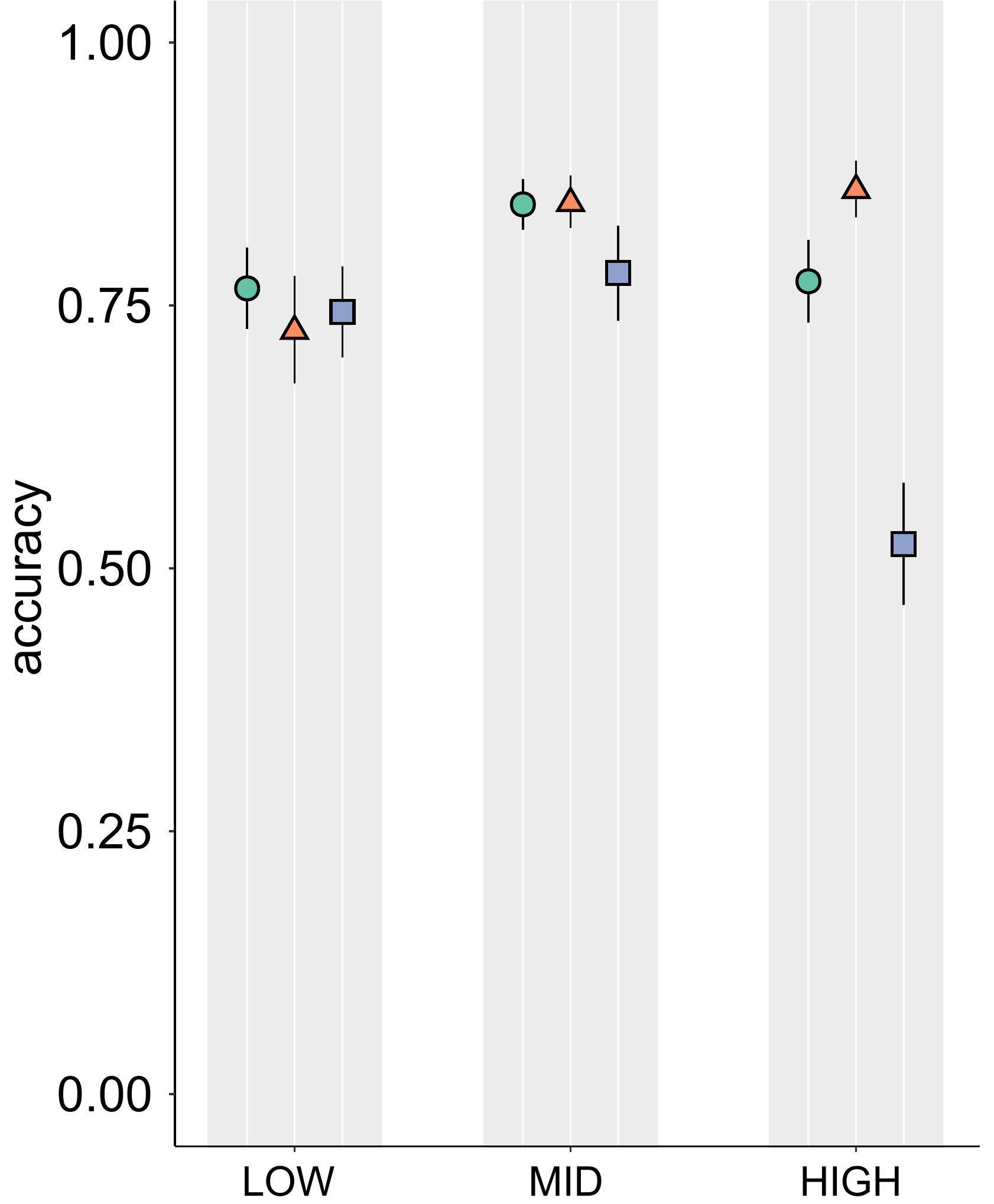}
    \caption{T3: Density Profile}
    \label{fig:t3_density}
\end{subfigure}
\hfill
\begin{subfigure}[t]{\var_width\textwidth}
      \centering
    \includegraphics[height=\varheight]{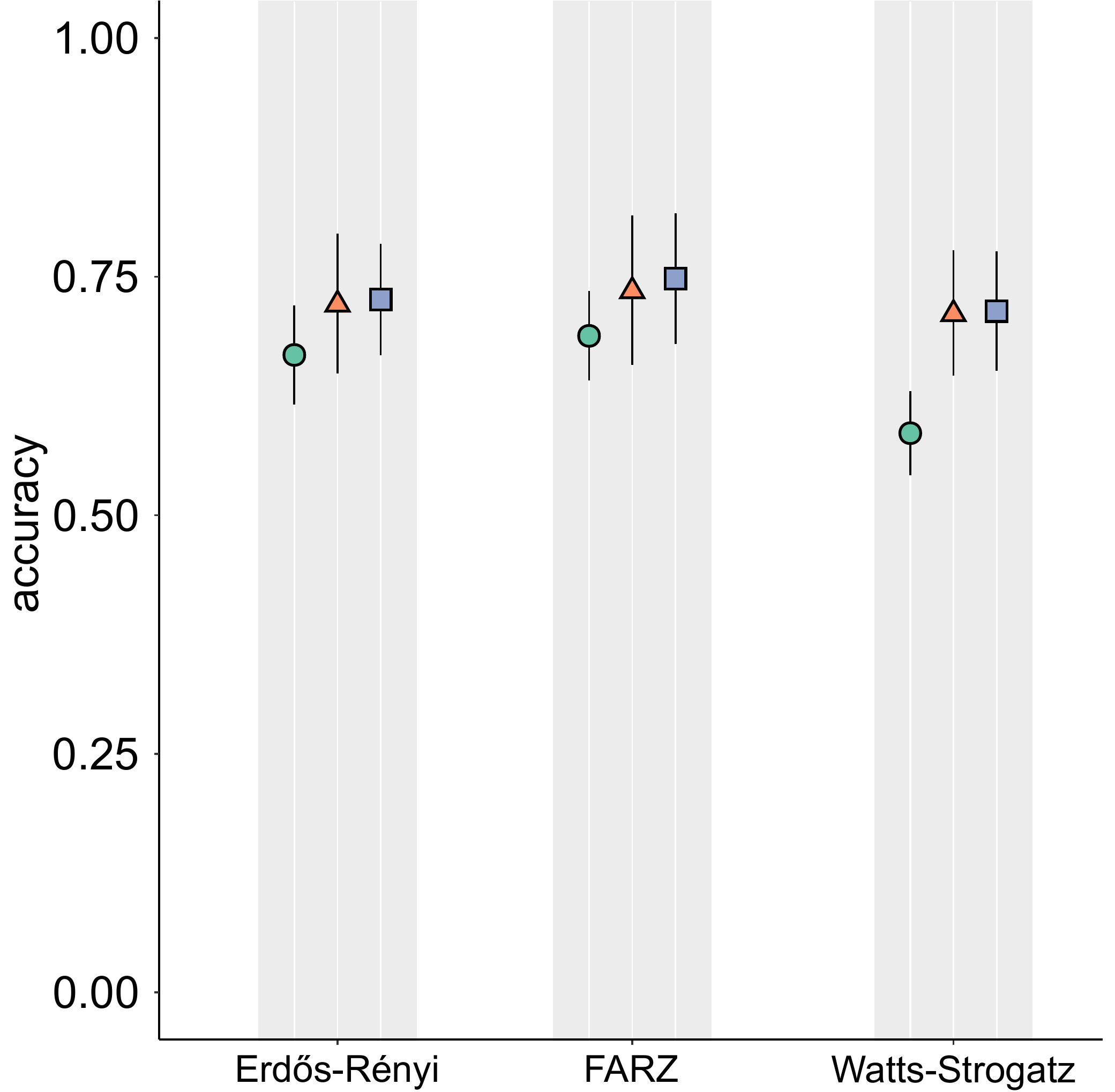}
    \caption{T4: Network Class Parameter}
    \label{fig:t4_classes}
\end{subfigure}
\hfill
\begin{subfigure}[t]{\var_width\textwidth}
      \centering
    \includegraphics[height=\varheight]{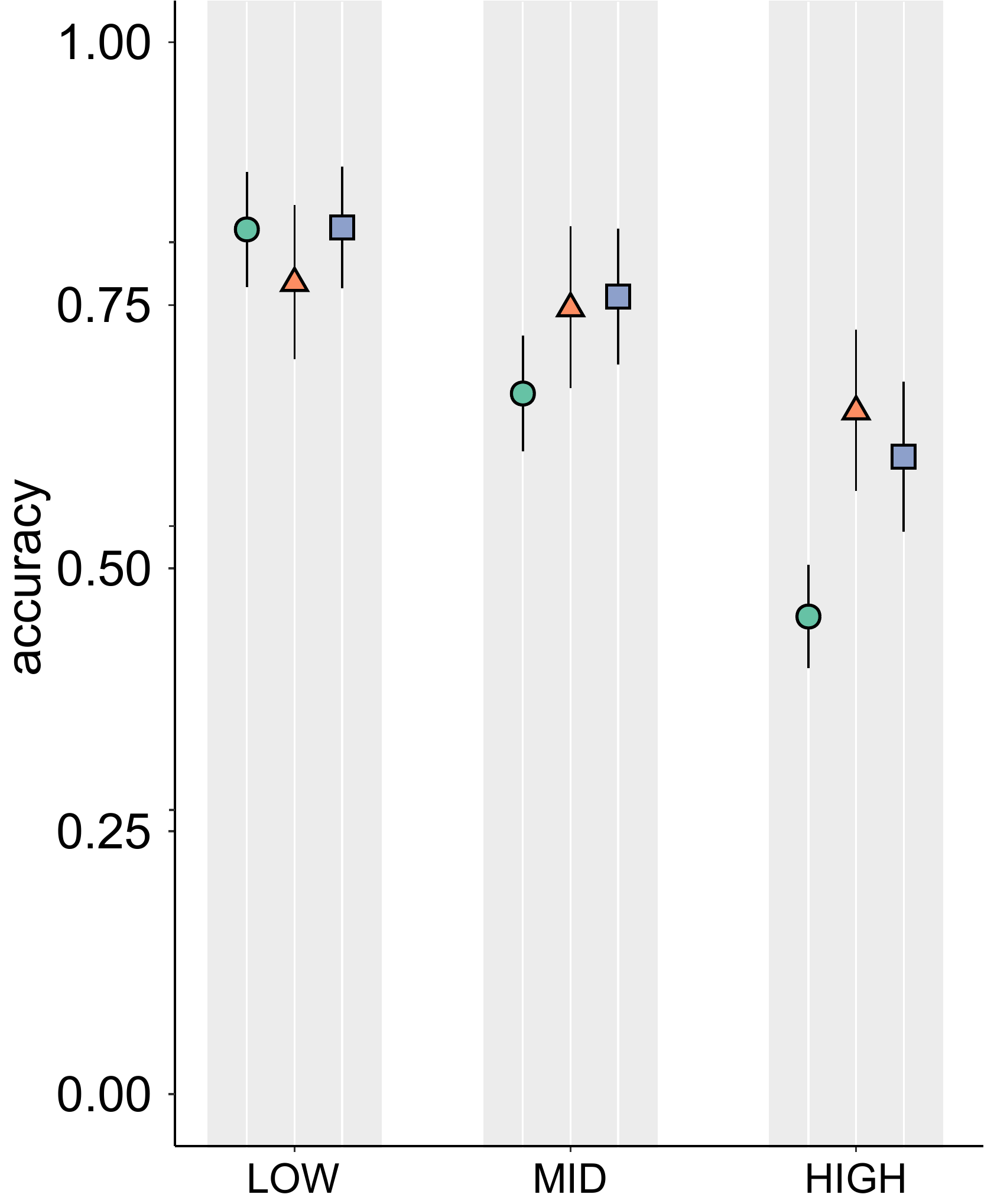}
    \caption{T4: Density Profile}
    \label{fig:t4_density}
\end{subfigure}

   \caption{The accuracy of each representation plotted with respect to different task parameters. The shapes represent the means, and the error bars are the 95\% confidence intervals. The accuracy is calculated by averaging first over the repetitions per task and then averaging over the mean per participant with respect to the considered parameter. }\vspace{-2ex}
   \label{fig:parameters}
\end{figure*}

\begin{table}[t]
\caption{Results of the overall statistical evaluation for the Kruskal-Wallis and pairwise Wilcoxon tests. p-values for multiple comparisons were Bonferroni corrected. We also report the confidence intervals of the mean difference \cite{cumming2005inference,cumming2014}.}
\label{tab:statistics}
\centering
 \resizebox{1.0\columnwidth}{!}{%
 \renewcommand{\arraystretch}{1.25}
\begin{tabular}{p{0.38\columnwidth}cccc}
\Xhline{3\arrayrulewidth}
\multicolumn{1}{l}{Hypothesis}  &   Vis Pair      & $\varB{p-value}$ & $M_{diff}$ & $CI_{diff}$ \\ \Xhline{3\arrayrulewidth}
\multirow{3}{*}{\parbox{0.4\columnwidth}{ \textbf{T1}: $H_1$(\ding{51})~ \\ $\chi^{2}(2)= 96.357$$, p<0.001$}}  
& \textbf{\BP} - \textbf{\NL} & \textless 0.001       & 0.33       & {[}0.28, 0.38{]}        \\ 
& \textbf{\AM} - \textbf{\BP} & 0.0031       & 0.05       & {[}0.02, 0.09{]}        \\ 
& \textbf{\AM} - \textbf{\NL} & \textless 0.001       & 0.38       & {[}0.33, 0.43{]}       \\ \hline
                     
\multirow{3}{*}{\parbox{0.4\columnwidth}{\textbf{T2}: $H_2$ (\ding{55})~ \\$\chi^{2}(2)= 42.342$$, p < 0.001$}}  
& \textbf{\NL} - \textbf{\BP} & \textless 0.001       & 0.15       & {[}0.10, 0.20{]}        \\
& \textbf{\AM} - \textbf{\BP} & \textless 0.001       & 0.11       & {[}0.06, 0.16{]}        \\
& \textbf{\NL} - \textbf{\AM} & 0.23       & 0.04       & {[}-0.01, 0.08{]}        \\ \hline
                     
\multirow{3}{*}{\parbox{0.4\columnwidth}{\textbf{T3}: $H_3(0)$ (\ding{55})~ \\$\chi^{2}(2)= 51.782$   $, p\textless{}0.001$}}  
& \textbf{\NL} - \textbf{\BP} & \textless 0.001       & 0.13       & {[}0.08, 0.17{]}        \\
& \textbf{\AM} - \textbf{\BP} & \textless 0.001       & 0.14       & {[}0.10, 0.19{]}        \\
& \textbf{\AM} - \textbf{\NL} & 0.21       & 0.02       & {[}-0.02, 0.05{]}     \\ \hline

\multirow{3}{*}{\parbox{0.4\columnwidth}{\textbf{T4}: $H_4$ (\partiallyConf )~  \\$\chi^{2}(2)= 19.54$$, p<0.001$}}  
&\textbf{\BP} - \textbf{\NL} & \textless 0.001       & 0.09       & {[}0.03, 0.15{]}        \\
&\textbf{\BP} - \textbf{\AM} & 1       & 0.01       & {[}-0.07, 0.08{]}        \\
&\textbf{\AM} - \textbf{\NL} & \textless 0.001       & 0.08       & {[}0.02, 0.15{]}     \\ \hline

\multirow{2}{*}{\parbox{0.4\columnwidth}{\textbf{T5}: $H_5$ (\ding{55})~ \\$W = 1411$, $ \varB{p-value} = 0.264$}}
& \multirow{2}{*}{\textbf{\BP} - \textbf{\AM}} & \multirow{2}{*}{0.264}       & \multirow{2}{*}{0.04}       & \multirow{2}{*}{{[}-0.04, 0.12{]}}        
\\\\ \Xhline{3\arrayrulewidth}

\end{tabular}
}
\vspace{-2ex}
\end{table}

\paragraph{\textbf{T1}} 
As seen in Figure~\ref{fig:results-overview}, we can confirm $H_1$ (\ding{51}), see also~\autoref{tab:statistics}. For \textbf{\NL} ($M = 0.54, SD = 0.15, CI = [0.49, 0.58]$), the task was solved less accurately compared to \textbf{\AM} ($M=0.92,SD=0.09,CI=[0.89, 0.94]$) and \textbf{\BP} ($M = 0.86, SD =$0.09$, CI = [0.84, 0.89]$). 
We also looked at the effect of network class and density pairing as seen in~\autoref{fig:t1_class} and~\autoref{fig:t1_density}. \textbf{\NL} seems to perform worst on Erdős-Rényi, while \textbf{\BP} seems best for Barabási. Further, low-low density pairings seem considerably worse than high-high density pairings for \textbf{\NL}.

\paragraph{\textbf{T2}} 
We cannot confirm $H_2$ (\ding{55}). \textbf{\BP} ($M = 0.66, SD = 0.13, CI = [0.62,0.69]$) performs worse than \textbf{\AM} ($M = 0.77, SD = 0.12, CI = [0.74,0.80]$), and there is a small difference for \textbf{\NL} ($M = 0.81, SD = 0.11, CI = [0.78,0.84]$) and \textbf{\AM}. 
\autoref{fig:t2_cluster} shows that \textbf{\NL} works best with 3 or 4 clusters, compared to the other representations. \textbf{\BP} performs worst for low-density profiles (\autoref{fig:t2_density}).

\paragraph{\textbf{T3}} 
We can reject $H_3(0)$ (\ding{55}), as we see an effect of network representation. \textbf{\BP} ($M = 0.76, SD = 0.12, CI = [0.72,0.79]$) leads to worse accuracy compared to AM ($M = 0.90, SD = 0.10, CI = [0.87,0.93]$) and \textbf{\NL} ($M = 0.88, SD = 0.08, CI = [0.86,0.91]$). 
\textbf{\BP} performs worse than \textbf{\AM} and \textbf{\NL} for high-density profiles (\autoref{fig:t3_density}). Similarly, \textbf{\BP} performs worse than \textbf{\NL} and \textbf{\AM}, when looking at network classes. It seems particularly bad for Barabási 
(\autoref{fig:t3_cluster}). For the delta parameter $\delta_d$, we could not see an effect.

\paragraph{\textbf{T4}} 
We can only partially confirm $H_4$ (\partiallyConf). \textbf{\AM} ($M = 0.80, SD = 0.21, CI = [0.74,0.86]$) is better than \textbf{\NL} ($M = 0.72, SD = 0.11, CI = [0.69,0.75]$), but on par with \textbf{\BP} ($M = 0.81, SD = 0.17,CI = [0.76,0.86]$). The results for the network class parameters reflect the overall trend and show that the task in the \textbf{\NL} condition was solved less accurately for Watts-Strogatz compared to \textbf{\AM} and \textbf{\BP} (\autoref{fig:t4_classes}). Density-wise, \textbf{\NL} performs worst on high-density profiles (\autoref{fig:t4_density}).

\paragraph{\textbf{T5}} 
As reflected in~\autoref{tab:statistics}, we did not find a significant effect of the representation for T5, so we cannot accept $H_5$ (\ding{55}). As there was no main effect, we also did not look deeper into the effect network class parameters might have.

\begin{figure}[t]
    \centering
    \includegraphics[width=\columnwidth]{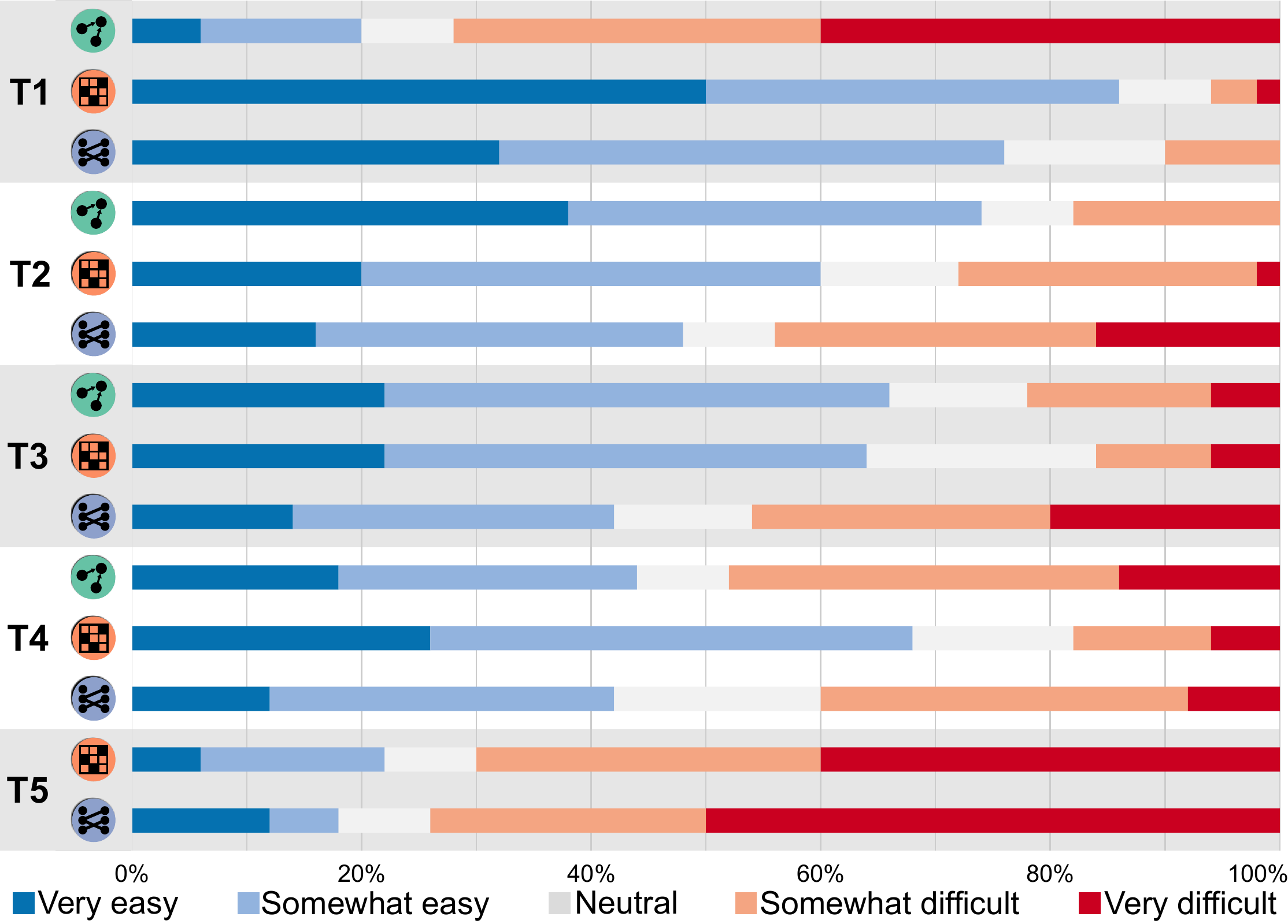}
    \caption{Subjective assessment of task difficulty.}\vspace{-2ex}
    \label{fig:task_difficulty}
\end{figure}
\section{Discussion}
Based on our results, we discuss the individual tasks, limitations, and finally, formulate guidelines for the use of the visualizations.

\subsection{Task Performance}

The results provide valuable information about how well the individual visualizations were suited to solve the tasks and what strategies were often applied by the participants. In this section, we discuss the performance results and shed light on the participants' subjective assessment of the task difficulty, along with their commentary feedback.

\paragraph{\textbf{T1}}

Both \textbf{\AM} and \textbf{\BP} are able to maintain the visual characteristics of each network class under varying densities. This is especially true when the network is sparse and the reference images are dense or vice versa (i.e., low-high density pairing). \textbf{\NL} seems to suffer when the given image or the reference images are of low density (\autoref{fig:t1_density}). 
While both \textbf{\AM} and \textbf{\BP} provide high accuracy, the results show that the classes seem to look more distinctive in \textbf{\AM}, except for Barabási networks (\autoref{fig:t1_class}). We assume that this is due to the low visibility of links in \textbf{\AM} when the network density is low, which might impair the recognition of class characteristics. In contrast, for \textbf{\BP}, such characteristics would be amplified on low-density networks.
From the qualitative feedback, we could identify that for \textbf{\NL}, density comparison was mentioned frequently: \textit{``... I tried to focus on density, or where many connections came together and then the overall shape.''} \textit{``I ... paid attention to how the nodes and edges were placed and tried to extrapolate how the network would look at different stages of denseness...''} The terms \textit{difficult} and \textit{hard} were mentioned more often than for the other techniques.
Overall, \textbf{\AM} and \textbf{\BP} were considered easier to compare: \textit{``(about \textbf{\AM})...I matched the pattern I saw, regardless of density...''}; \textit{``(about \textbf{\BP}) I just looked at the overall pattern. It seemed like it was either a diagonal,  an x-shape, bands, or solid.''} This impression is also reflected in task difficulty (Figure~\ref{fig:task_difficulty}).

\paragraph{\textbf{T2}}
We hypothesized that \textbf{\NL} and \textbf{\BP} will be more accurate than \textbf{\AM}. The results could not support our hypothesis. We noticed that denser networks increase the accuracy, regardless of the representation (\autoref{fig:t2_density}). However, the results show that \textbf{\BP} is more affected than \textbf{\NL} and \textbf{\AM} when the network density is low. As it appears, the visibility of links plays a less important role than the proximity of nodes, when it comes to detecting clusters in sparse networks. This makes \textbf{\BP} and \textbf{\AM} less suited than \textbf{\NL} because both representations rely on link visibility to reveal cluster structure when the nodes are ordered in a proper way (\autoref{fig:t2_density}). 
However, the results suggest that the overdrawing in \textbf{\BP} might have impaired the recognition of clusters, in comparison to \textbf{\AM}. This is also reflected in the participants' comments. It was frequently mentioned that the task became hard to solve with \textbf{\BP} for sparse networks. \textit{``My strategy was to look for the darker lines as they seemed to be a cut-off point for one cluster. This was not always evident or obvious in the more sparse clusters.''}; \textit{``I tried to count how many bands there were. It was a little hard for sparser networks.''} A similar approach was mentioned for \textbf{\AM}: \textit{``I looked for the spots that were most dense and counted them.''} However, difficulties with sparse networks were less frequently mentioned than for \textbf{\BP}. For \textbf{\NL}, many participants reported that the task was easy and they did not provide further details on their strategies.
By investigating the influence of the number of clusters in relation to the accuracy, we found that participants tended to underestimate the number of clusters in the network more than overestimate them (\autoref{fig:t2_cluster}). Nevertheless, both effects occurred with low-density and mid-density profiles. This is understandable since the network size is constant while the number of clusters $k$ varies.

\paragraph{\textbf{T3}}
In general, \textbf{\NL} and \textbf{\AM} are significantly more accurate than \textbf{\BP}. The results suggest that \textbf{\BP} has the least sensitivity to density changes. 
The density plot (\autoref{fig:t3_density}) shows that \textbf{\NL} and \textbf{\BP} are slightly better than \textbf{\AM} at low densities, whereas \textbf{\AM} outperforms both at high densities. 
However, \textbf{\BP} seems to be impacted at high densities. 
\textbf{\BP}, and to a lesser degree \textbf{\NL}, rely on transparency to reduce visual clutter. As such, the denser the network, the lighter and smoother the lines will appear. 
For a non-trained eye, that might be misleading. The participants' comments support this: \textit{``(about \textbf{\BP}) I tried to focus more on the white space and darkness of the images overall.''}; \textit{``For the ones that had the darker images, I looked at the darkest area and checked to see how smooth it was...''} 
In contrast, \textbf{\AM} relies on the \textit{density}, which increases the overall contrast of the image, making the comparison task more intuitive. This is also reflected in the participants' comments: \textit{``(about \textbf{\AM}) I chose the square that was more dense or darker.''}
With respect to the network class, \autoref{fig:t3_cluster} shows that for \textbf{\AM} and \textbf{\NL}, Erdős-Rényi obtained the most accurate results among the four classes, which is consistent with what we expected. However, the performance did not degrade that much for FARZ and Watts-Strogatz. This suggests that changes in the overall structure of the network did not impair the participants from finding which network has more connections. 
For \textbf{\BP}, the results in~\autoref{fig:t3_cluster} suggest that the task was harder to solve on Barabási in comparison to the other classes. One explanation could be that the participants focused more on the top part of the diagram where the main hubs are located and neglected other parts of the network. 

\paragraph{\textbf{T4}}

\textbf{\NL} is not well suited to solve this task, especially when the network density is high. Contrary to our expectation, the performance of \textbf{\BP} seems to be on par with \textbf{\AM}, suggesting that visual clutter did not impair the participants from estimating the node degree (\autoref{fig:t4_density}). Looking at the participants' comments revealed slightly different strategies to solve the task on \textbf{\BP} versus \textbf{\AM}. For \textbf{\BP}, the words "estimating" or "squinting" was often mentioned:  \textit{``...some squinting on very dense clusters. A good bit of counting on more sparse graphs where you could make out individual lines...''};\textit{``I tried to estimate how many were on each side. It helped to blur my vision slightly and look at the density of the lines.''}
For \textbf{\AM}, participants mentioned: \textit{``I counted the black squares unless they were a lot...''};\textit{``I counted the empty white squares...''}

\paragraph{\textbf{T5}}

Although the results show that \textbf{\BP} is more accurate than \textbf{\AM}, the difference was not significant (\autoref{fig:results-overview}). While there are few participants who managed to obtain 100\% accuracy, the majority found this task hard to solve, regardless of the representation (\autoref{fig:task_difficulty}). 
We assume that most participants did not spend enough time solving this task. The data revealed that, on average, participants spent about 30 sec per trial. Only 9 participants in \textbf{\BP} and 6 participants in \textbf{\AM} managed to obtain an accuracy of more than ~83\%. The data revealed that these participants tended to spend one minute on average for each trial. Looking at the comments reveals some strategies. \textbf{\BP} was investigated to identify missing and existing connections: \textit{``I checked the points that had no connections and also those points that made multiple connections to make it easier to identify the correct option.''}; \textit{``Check how many solo nodes there were and then compare to the diagram. Also checking nodes that had a lot of connections coming in...''} A similar behavior was mentioned for \textbf{\AM}: \textit{``I tried to match up the empty rows/columns with how many outliers there were, and if necessary counted the largest amount of connections...''} A short attention span is a common pitfall of crowd-sourced evaluation~\cite{Archambault2017}. We believe that the performance of this task might have been better if it had been done in a controlled-lab setup. 

With respect to the comparison between \textbf{\NL} and \textbf{\AM}, our findings are consistent with the previous work with respect to task T1~\cite{Yoghourdjian2018}, T2~\cite{Okoe2018,Okoe2019,Berardi2013efficacy}, and T3~\cite{Okoe2018,Okoe2019, Yoghourdjian2018}, despite the fact that authors evaluated smaller networks. With respect to task T4, our findings partially match the previous work~\cite{Hlawatsch2014Adjacency, Keller2006} in low-density profiles. However, we found that \textbf{\AM} performs significantly better than \textbf{\NL} in dense networks.

\subsection{Limitations}
The results of our crowd-sourced study provide new insights for the visualization of network data. Regarding the number of participants, we are on par with other work~\cite{Okoe2018,Radu2014HowtoDisplay}.
However, we identified some limitations for the interpretation of the results.

\paragraph{Crowd-sourced Evaluation}
Crowd-sourced evaluation has the benefit of large sample sizes but also comes with problems~\cite{Archambault2017}. Our pilot testing showed that the study needs about 45 min for a novice participant (a participant who has no prior knowledge of network visualization) to complete. To ensure the validity of our results, we set our filtering criteria based on the time spent to complete the study but also based on the successful responses to the attention checks we put throughout the study. We originally recruited 240 unique online users to take our experiment, but only 150 (62\%) passed our filtering criteria. This shows that some of the participants were not paying attention during the study, which had to be handled later by the experimenters.

\paragraph{Network Size}

In our study, we fixed the network size parameter to not increase the evaluations' complexity. For the overview tasks, we chose size 500 as it appeared to be the fairest for the three representations, considering that we do not include interactive features in the evaluation. Based on our experience, we argue that the study findings would still hold for \textbf{\NL} and \textbf{\BP} at a network size of 1000. But it might be different for \textbf{\AM}. It is an open question how each representation would scale as the network grows even beyond 1000 nodes.

\paragraph{Network Representation}

In our experiment, we tried to be fair in selecting the layout methods and clutter-reduction techniques. This can explain why we did not include edge bundling techniques in the evaluation. While edge bundling would certainly be beneficial for \textbf{\NL} and \textbf{\BP}, it does not have a direct mapping in the \textbf{\AM} representation, and the question of how edge bundling influences the appearance of network structures would require more investigation that we deemed beyond the scope of the study. For layout methods,  we used the \textit{d3-force}~\cite{d3-force:online} algorithm to lay out \textbf{\NL} and hierarchical clustering~\cite{hclustfu46:online} to order the nodes in \textbf{\AM} and \textbf{\BP} representations. Our selection of these methods was driven by their wide adoption in the community~\cite{bostock2011d3,Behrisch2016Matrix} and the availability of their implementations in many languages/tools. Nevertheless, there is still room for improvement to ensure the fairest comparison between the different representations~\cite{ModularityNoack2009}.

\paragraph{Selected Tasks}

The tasks evaluated in this study are not meant to be an exclusive list. During the data study, we experimented with eight different tasks. We decided to only include five tasks to not extend the study beyond a realistic length. The other three tasks that we did not include are: \textit{identifying common neighbors between two nodes}, \textit{looking up a node}, and \textit{finding a path between two nodes}. In contrast to the overview tasks we included in the study, these three are detailed network tasks. 
While these tasks are certainly valid candidates for further evaluation, we decided to focus on network overview tasks, since they were not often considered by other 
studies~\cite{YOGHOURDJIAN2018264}.

\subsection{Guidelines}

In this work, we evaluated three overview tasks (T1 -- T3) and two detailed tasks (T4 and T5) for directed networks. Depending on the task at hand, one representation might perform better than others. For tasks that involve comparing the overall network structure, \textbf{\AM} and \textbf{\BP} might be better candidates than \textbf{\NL} . Both representations are versatile enough to depict the various structural properties and maintain these properties under varying densities. In contrast, when the task involves identifying network clusters or sub-communities, \textbf{\NL} seems to be superior to both representations, since the cluster information is double encoded by the proximity and connectivity between the nodes.
Due to the overdrawing problem, \textbf{\BP} does not seem to be a good choice when the task is about estimating overall density or detecting subtle changes in connectivity. In such situations, \textbf{\AM} provides the most accurate estimation of the true density within the network. Finally, for tasks that are based on node attributes, \textbf{\AM} and \textbf{\BP} offer a better overview, especially in the context of directed networks, as both representations have two separate axes for encoding source and target nodes. Overall, \textbf{\AM} seems to be the most reliable among the three representations across the different tasks.

\section{Conclusion}
We presented a comparative user study about node-link diagrams, adjacency matrices, and bipartite layouts for the visualization of network data. By assessing the performance in five different tasks, we derived a set of guidelines for using the techniques with respect to the task and network properties at hand. Overall, this study shows the strengths and limitations of established graph visualization techniques and provides insights into when a bipartite layout might help interpret the data. In the future, we want to investigate solution strategies for the tasks in detail. Eye tracking methodology might help understand how people read the visualizations and provide more information about conscious and subconscious processes than post-experimental self-reports. Further, we plan to extend the catalog of tasks to derive a more comprehensive set of guidelines for using the visualizations.

\acknowledgments{
This work is supported by the Deutsche Forschungsgemeinschaft (DFG, German Research Foundation) under Germany's Excellence Strategy -- EXC 2120/1 -- 390831618 and -- Project ID 251654672 -- TRR 161.}

\bibliographystyle{abbrv-doi-hyperref-narrow}

\bibliography{template}
\end{document}